\documentclass[reprint,twocolumn,superscriptaddress,showkeys,
nofootinbib,notitlepage,amsmath,amssymb,floatfix]{revtex4-1}

\pdfoutput=1
\usepackage{latexsym,amsmath,amssymb,lmodern,float,url}
\usepackage{natbib}
\usepackage{color}
\usepackage{multirow}
\usepackage{bm}
\usepackage{array}
\usepackage{graphicx}

\def\de{\delta^{\vphantom{1}}}
\def\bde{{\bar\delta}}
\def\qq{{q\bar q}}
\def\QQ{{Q\bar Q}}

\def\h3{{\displaystyle{\frac 3 2}}}

\def\Mb0{{\overline{M}_0}}
\def\kqQb{{\kappa_{qb}}}
\def\kqQcs{{\kappa_{sc}}}
\def\sqb{{\epsilon_{qb}}}
\def\ssc{{\epsilon_{sc}}}

\begin{document}
\title{Spectrum of the Hidden-Bottom and the Hidden-Charm/Strange
Exotics in the Dynamical Diquark Model}
\author{Jesse F. Giron}
\email{jfgiron@asu.edu}
\author{Richard F. Lebed}
\email{Richard.Lebed@asu.edu}
\affiliation{Department of Physics, Arizona State University, Tempe,
AZ 85287, USA}
\date{May, 2020}

\begin{abstract}
The lightest hidden-bottom tetraquarks in the dynamical diquark model
fill an $S$-wave multiplet consisting of 12 isomultiplets.  We predict
their masses and dominant bottomonium decay channels using a simple
3-parameter Hamiltonian that captures the core fine-structure features
of the model, including isospin dependence.  The only experimental
inputs needed are the corresponding observables for $Z_b(10610)$ and
$Z_b(10650)$.  The mass of $X_b$, the bottom analogue to $X(3872)$, is
highly constrained in this scheme.  In addition, using
lattice-calculated potentials we predict the location of the center of
mass of the $P$-wave multiplet and find that $Y(10860)$ fits well but
the newly discovered $Y(10750)$ does not, more plausibly being a
$D$-wave bottomonium state.  Using similar methods, we also examine
the lowest $S$-wave multiplet of 6 $c\bar c s\bar s$ states, assuming
as in earlier work that $X(3915)$ and $Y(4140)$ are members, and
predict the masses and dominant charmonium decay modes of the other
states.  We again use lattice potentials to compute the centers of
mass of higher multiplets, and find them to be compatible with the
masses of $Y(4626)$ ($1P$) and $X(4700)$ ($2S$), respectively.
\end{abstract}

\maketitle

\section{Introduction}

The modern study of hadrons that manifest exotic va\-lence-quark
content has produced numerous surprises in both experiment and theory,
as reviewed in Refs.~\cite{Lebed:2016hpi,Chen:2016qju,Hosaka:2016pey,
Esposito:2016noz,Guo:2017jvc,Ali:2017jda,Olsen:2017bmm,
Karliner:2017qhf,Yuan:2018inv,Liu:2019zoy,Brambilla:2019esw}.  As of
this writing, more than 40 candidates have been observed in the
heavy-quark sector.  However, the fundamental organizing principle
underlying their spectroscopy has proved elusive, unlike the clear
structure derived from quark-potential models in the conventional
$c\bar c$ and $b\bar b$ sectors~\cite{Tanabashi:2018oca}.

For instance, one may attempt to model multiquark exotics using the
original molecular picture of two conventional hadrons bound via
light-meson ({\it e.g.}, $\pi$)
exchange~\cite{Voloshin:1976ap,Tornqvist:1993ng}.  This approach can
provide some guidance regarding which thresholds might be expected to
support a molecule~\cite{Cleven:2015era,Karliner:2015ina}.  However,
hadronic molecules lack a regularly spaced spectrum because the
pattern of mass splittings among the light and heavy-light hadrons
acting as their constituents is itself nontrivial, being obscured by
the specifics of strong-interaction dynamics.  In addition, a
composite state of a given width cannot form if either constituent
hadron has a larger width, and it remains unclear whether molecular
formation is limited to the case in which the constituents are in a
relative $S$ wave.  Indeed, calculating the detailed properties of
hadronic molecules appears to require the careful consideration of a
variety of near-threshold effects such as cusps and rescattering
diagrams~\cite{Guo:2017jvc}.

The $J^{PC} \! = \! 1^{++}$ $X(3872)$, the first heavy-quark exotic
state discovered~\cite{Choi:2003ue}, is touted as the {\it ne plus
ultra\/} of hadronic molecules, and no reasonable researcher can deny
that the absurdly small splitting $m_{X(3872)} \! - \! m_{D^0} \! - \!
m_{D^{*0}} \! = \! +0.01 \! \pm \! 0.18$~MeV indicates the controlling
influence of the state $D^0 \bar D^{*0}$ (charge conjugates
understood) over the nature of the resonance.  And yet, the very
proximity of $X(3872)$ to threshold indicates that it is almost
certainly not a ``traditional'' molecule of the type described above,
but rather its precise mass eigenvalue relies in an intrinsic way upon
threshold effects.  Several observed features of $X(3872)$ point to a
complicated structure; for example, its substantial collider prompt
production rate suggests that $X(3872)$ possesses a tightly bound
component, but the suppression of this rate with increasing
charged-particle multiplicity in $pp$ collisions as compared to that
of $\psi(2S)$~\cite{Durham:2020zuw} suggests that $X(3872)$ may more
easily dissociate in a dense particle environment, as one expects for
a molecule.  A typical resolution of this conundrum is to suppose that
the $1^{++}$ conventional charmonium state $\chi_{c1}(2P)$, predicted
by potential models to lie around 3925~MeV~\cite{Barnes:2005pb} but
conspicuously absent from the data, mixes to a significant degree with
a $D^0 \bar D^{*0}$ state to form the physical $X(3872)$.

The $\chi_{c1}(2P)$ is not the only example of a tightly bound state
that can mix with $X(3872)$.  Diquark models also produce a single
isoscalar $1^{++}$ tetraquark state as one of their lowest
hidden-charm excitations, appearing in the color-attractive
arrangement $(cq)_{\bf \bar 3}(\bar c \bar q)^{\vphantom\dagger}_{\bf
3}$~\cite{Maiani:2004vq}.  Typical diquark $\de \! \equiv \! (cq)_{\bf
\bar 3}$ masses of $\sim \! 1.9$~GeV naturally produce such a $1^{++}$
$\de$-$\bde$ state in the vicinity of $3.9$~GeV~\cite{Giron:2019bcs}.
$X(3872)$ might actually, in the end, prove to be a perfect storm of a
$D^0 \bar D^{*0}$ molecular state enhanced by threshold effects,
mixing with the otherwise isolated conventional charmonium
$\chi_{c1}(2P)$ state and the lowest-lying isoscalar $1^{++}$
$\de$-$\bde$ state.

The variant diquark model used in this work is the so-called
``dynamical'' diquark model, which was
developed~\cite{Brodsky:2014xia} to address the issue of how
$\de$-$\bde$ states persist long enough to be observed, rather than
their quarks instantly recombining through the more strongly
attractive ${\bf 3} \otimes \bar{\bf 3} \! \to \! {\bf 1}$ color
coupling into meson pairs.  The physical picture has two components:
First, a heavy quark $Q$ must be created in closer proximity to a
quark $q$ than to an antiquark $\bar q$, and form a somewhat compact
diquark quasiparticle $\de \! \equiv \! (Qq)_{\bar{\bf 3}}$, and vice
versa for $\bde$; and second, the large energy release of the
production process drives apart the $\de$-$\bde$ pair before
recombination into a meson pair can occur, creating an observable
resonance.\footnote{Equivalently, the full four-quark wave function
has a large overlap with two-meson states when its $\de$,$\bde$
components have a small relative momentum, and a large overlap with an
idealized $\de$-$\bde$ state (and a suppressed overlap with two-meson
states) when this relative momentum is large.}  A similar mechanism
using color-triplet {\it triquarks\/} extends the picture to
pentaquark formation~\cite{Lebed:2015tna}.

This physical picture was developed into a predictive
model~\cite{Lebed:2017min} by describing the color flux tube that
connects the separating $\de$-$\bde$ pair using the language of
potentials in the Born-Oppenheimer (BO) approximation.  These
potentials are the same ones appearing in QCD lattice gauge-theory
simulations of heavy-quark hybrid mesons~\cite{Juge:1997nc,
Juge:1999ie,Juge:2002br,Morningstar:2019,Capitani:2018rox}, so they
may be applied directly to obtain numerical results for the
$\de$-$\bde$ spectrum~\cite{Giron:2019bcs}, since both systems involve
glue connecting heavy color ${\bf 3}$ and $\bar{\bf 3}$ sources.  The
lowest BO multiplets are all found numerically to lie in the
$\Sigma^+_g$ potential\footnote{A full definition of the standard BO
potential notation is presented in Ref.~\cite{Lebed:2017min}.  The
specific case $\Sigma^+_g$ means that the projection of angular
momentum along the axis connecting the heavy sources has eigenvalue
0, and that the light degrees of freedom are symmetric under two
reflections: through a plane perpendicular to and bisecting this
axis, and under the $CP$ inversion of the light degrees of freedom
(using the midpoint of the heavy sources as the origin).}, and in
order of increasing mass are $1S$, $1P$, $2S$, $1D$, and $2P$.  The
parity of all states in each $\Sigma^+_g$ multiplet is simply given
by $(-1)^L$.

As first proposed in Ref.~\cite{Maiani:2014aja}, the dominant
spin-spin couplings in the $\de$-$\bde$ states (as supported by
comparison to observation) appear to be the ones {\em within\/} each
of $\de$ and $\bde$.  The strength of this coupling is denoted by
$\kappa_{qQ}$, where $Q$ refers to the heavy quark and $q$ the light
quark in $\de$.  The dominance of these particular spin couplings
arises naturally if $\de$, $\bde$ are more compact than the full
exotic state in which they appear.  Furthermore, the near-universal
prediction that spin-singlet couplings within diquarks are more
attractive than spin-triplet couplings leads to the expectation that
$\kappa_{qQ} \! > \! 0$.  A detailed numerical examination of the
effect of including a finite diquark size is one of the primary
thrusts of Ref.~\cite{Giron:2019cfc}; there it is found that the
calculated state masses are remarkably stable as long as the diquark
wave functions no longer significantly overlap when the distance $R$
between their centers exceeds the critical value of 0.8~fm.  In other
words, the diquarks may have radii as large as
$R/2 \! \simeq \! 0.4$~fm and still be considered compact for the
purpose of the model.  Indeed, Ref.~\cite{Giron:2019cfc} also found
that observation [specifically, the experimental absence of a charged
partner to $X(3872)$] does not support the dominant isospin
dependence in the $\de$-$\bde$ state being one that couples to
diquarks as truly pointlike objects, but the model works quite well
when the dominant isospin dependence is instead taken to couple only
to the light quarks within $\de$ and $\bde$.  (And of course, isospin
exchange is irrelevant for $c\bar{c}s\bar{s}$ states.)

The mass spectrum and preferred heavy-quark spin-eigenstate decay
modes of the 12 isomultiplets (6 isosinglets and 6 isotriplets)
comprising the $c\bar c q \bar q^\prime$ $\Sigma^+_g(1S)$ multiplet
($q,q^\prime \in \{ u, d \}$) was studied in
Ref.~\cite{Giron:2019cfc}.  This was the first work to differentiate
$I \! = \! 0$ and $I \! = \! 1$ states in a diquark model.  The model
naturally produces scenarios in which $X(3872)$ is the lightest
member; and of the two $I \! = \! 1$, $J^{PC} \! = \!  1^{+-}$
states, the $Z_c(3900)$ naturally decays to $J/\psi$ and the
$Z_c(4020)$ to $h_c$, as is observed.  The simplest model uses a
3-parameter Hamiltonian: a common multiplet mass, an internal
diquark-spin coupling, and a long-distance isospin-dependent coupling
between the light quark $q$ in $\de$ and light antiquark
$\bar q^\prime$ in $\bde$.  The corresponding analysis of the 28
states of the negative-parity $c\bar c q \bar q^\prime$
$\Sigma^+_g(1P)$ multiplet, which includes precisely 4 $Y$ states
($J^{PC} \! = \!  1^{--}$), was performed in
Ref.~\cite{Giron:2020fvd}.  In this case, the simplest model has 5
parameters, including now spin-orbit and tensor terms.  An earlier
diquark analysis using a similar Hamiltonian but not including isospin
appears in Ref.~\cite{Ali:2017wsf}.

In this paper we extend the study of the dynamical diquark model to
the $\Sigma^+_g(1S)$ multiplet in the hidden-bottom ($b\bar b q \bar
q^\prime$) sector (again, 12 isomultiplets) and the hidden-charm,
hidden-strange ($c\bar c s\bar s$) sector (6 states).  Remarkably,
using only the well-known $Z_b(10610)$ and $Z_b(10650)$ states---often
themselves identified as $B\bar B^*$ and $B^* \! \bar B^*$ molecules,
respectively---and rough information from their $\Upsilon$ and $h_b$
branching ratios, one can predict masses of the remaining 10 states
and their preferred heavy-quark decay channels with surprising
accuracy.  The analysis of the $c\bar c s\bar s$ states builds upon
that of Ref.~\cite{Lebed:2016yvr} [which assumes that $X(3915)$ and
$Y(4140)$ are $c\bar c s\bar s$ states] to reflect the current state
of data and to develop a better understanding of the underpinnings of
the model.  The negative-parity $\Sigma^+_g(1P)$ multiplet consists of
28 states for $b\bar b q \bar q^\prime$ and 14 states for $c\bar c
s\bar s$, but only a very small number of candidates have been
observed for each type; nevertheless, we use the approach of
Ref.~\cite{Giron:2019bcs} to predict the centers of mass of the
$\Sigma^+_g(1P)$ and $\Sigma^+_g(2S)$ multiplets, and find that most of
the candidates lie in the anticipated mass regions [the exception
being $Y(10750)$, which we argue to be a conventional bottomonium
state].

This paper is organized as follows: In Sec.~\ref{sec:Expt} we review
the current data on $b\bar b q\bar q^\prime$ and $c\bar c s\bar s$
candidates.  Section~\ref{sec:MassHamOp} reprises the analysis of
Ref.~\cite{Giron:2019cfc}, as applied to these sectors.  The naming
scheme for levels comprising the $\Sigma^+_g(1S)$ multiplets is
defined in Sec.~\ref{sec:MassExp}, and we present explicit expressions
for their masses in terms of the model parameters.  Numerical analysis
of states in the $b\bar b q\bar q^\prime$ and the $c\bar c s\bar s$
sectors appears in Section~\ref{sec:Results}, where both mass
eigenvalues and mixing parameters relevant to heavy-quark decay modes
are predicted.  We conclude in Sec.~\ref{sec:Concl}.

\section{Experimental Review}
\label{sec:Expt}

\subsection{The $b\bar{b}q\bar{q}^\prime$ Sector}
\label{sec:bbqq_ExptReview}
\begin{table*}[ht]
  \caption{All bottomoniumlike exotic-meson candidates catalogued by
  the Particle Data Group (PDG)~\cite{Tanabashi:2018oca}.  Also
  included is the recently observed
  $Y(10750)$~\cite{Abdesselam:2019gth}.  Both the particle name most
  commonly used in the literature and its label as given in the PDG
  are listed.  Only bottomonium decays are listed, and branching
  ratios are given where available.}
\label{tab:bbqqExpt}
\centering
\setlength{\extrarowheight}{1.5ex}
\begin{tabular}{cccccrl}
\hline\hline
    Particle
        & PDG label
            & $I^{G} \, J^{PC}$
                & Mass [MeV]
                    & Width [MeV]
                        & \multicolumn{2}{c}{Production and Decay} \\
			\hline
    $Z_b(10610)^{\pm}$ & $Z_b(10610)^\pm$
        & $1^+ \, 1^{+-}$
            & $10607.2\pm 2.0$
                & $18.4\pm 2.4$
                    & $ e^+e^-\to Z$; & $Z \to\left\{
		    \begin{array}{ll}
                    \Upsilon(1S)\pi^+\pi^- &
			\left( 5.4^{+1.9}_{-1.5}\right)\times 10^{-3}
			\\
                    \Upsilon(2S)\pi^+\pi^- &
			\left( 3.6^{+1.1}_{-0.8}\right) \; \% \\
                    \Upsilon(3S)\pi^+\pi^- &
			\left( 2.1^{+0.8}_{-0.6}\right) \; \% \\
                    h_b(1P)\pi^+\pi^- &
			\left( 3.5^{+1.2}_{-0.9}\right) \; \% \\
                    h_b(2P)\pi^+\pi^- &
			\left(4.7^{+1.7}_{-1.3}\right) \; \%
                    \end{array} \right. $ \\
    $Z_b(10610)^{0}$ & $Z_b(10610)^0$
        & $1^+ \, 1^{+-}$
            & $10609\pm 6$
                & $18.4\pm 2.4$
                    & $e^+ e^- \to Z$; & $Z \to\left\{
		    \begin{array}{l}
                    \Upsilon(2S)\pi^0\\
                    \Upsilon(3S)\pi^0\\                
                    \end{array} \right. $ \\
    $Z_b(10650)^{\pm}$ & $Z_b(10650)^\pm$
        & $1^+ \, 1^{+-}$
            & $10652.2 \pm 1.5 $
                & $11.5 \pm 2.2$
                    & $e^+ e^- \to Z$; & $Z \to\left\{
		    \begin{array}{ll}
 		    \Upsilon(1S)\pi^+\pi^- &
			\left(1.7^{+0.8}_{-0.6} \right)\times 10^{-3}
			\\
                    \Upsilon(2S)\pi^+\pi^- &
			\left(1.4^{+0.6}_{-0.4}\right)\;\%\\
                    \Upsilon(3S)\pi^+\pi^- &
			\left(1.6^{+0.7}_{-0.5} \right)\;\%\\
                     h_b(1P)\pi^+\pi^- &
			\left(8.4^{+2.9}_{-2.4} \right)\;\%\\
                     h_b(2P)\pi^+\pi^- & \left(15\pm 4\right)\;\%\\
		    \end{array} \right. $ \\
    $Y(10750)$ & $\Upsilon(10750)$
        & $0^- \, 1^{--}$
            & $10752.7^{+5.9}_{-6.0}$
                 & $35.5^{+18.0}_{-11.8}$
                    & $e^+ e^- \to \gamma Y$; & $Y \to \left\{
		    \begin{array}{ll}
                    \Upsilon(1S)\pi^+\pi^-\\
                    \Upsilon(2S)\pi^+\pi^-\\
                    \Upsilon(3S)\pi^+\pi^-\\
		    \end{array} \right. $ \\
    $Y(10860)$ & $\Upsilon(10860)$
        & $0^- \, 1^{--}$
            & $10889.9^{+3.2}_{-2.6}$
                & $51^{+6}_{-7}$
                    &  $e^+e^- \to \gamma Y$; & $Y \to \left\{
		    \begin{array}{ll}
                    \Upsilon(1S)\pi^+\pi^- &
			\left(5.3\pm 0.6\right)\times 10^{-3}\\
                    \Upsilon(2S)\pi^+\pi^- &
			\left(7.8\pm 1.3\right)\times 10^{-3}\\
                    \Upsilon(3S)\pi^+\pi^- & 
			\left(4.8^{+1.9}_{-1.7}\right)\times 10^{-3}\\
                    \Upsilon(1S)K^+K^- &
			\left(6.1\pm 1.8\right)\times 10^{-4}\\
                    h_b(1P)\pi^+\pi^- &
			\left(3.5^{+1.0}_{-1.3}\right)\times 10^{-3}\\
                    h_b(2P)\pi^+\pi^- &
			\left(5.7^{+1.7}_{-2.1}\right)\times 10^{-3}\\
                    \eta \Upsilon_J(1D) &
			\left(4.8\pm 1.1\right)\times 10^{-3}\\
                    \chi_{b1}(1P)\pi^+\pi^-\pi^0 &
			\left(1.85\pm 0.33\right)\times 10^{-3}\\
                    \chi_{b2}(1P)\pi^+\pi^-\pi^0 &
			\left(1.17\pm 0.30\right)\times 10^{-3}\\
                    \end{array} \right. $ \\ \hline\hline
\end{tabular}
\end{table*}

Of all hidden-bottom states thus far observed, only a handful are
exotic candidates, which are summarized in Table~\ref{tab:bbqqExpt}.
The most familiar examples are the $I \! = \! 1$, $J^{PC} \! = \!
1^{+-}$ states $Z_b(10610)$ and $Z_b(10650)$.  Their proximity to the
thresholds for $B \bar B^*$ ($10604.2 \pm 0.3$~MeV) and $B^* \! \bar
B^*$ ($10649.4 \pm 0.4$~MeV), respectively, suggests a natural
identification as molecular states.  These states also possess
hidden-charm analogues $Z_c(3900)$ and $Z_c(4020)$ that carry the same
quantum numbers, which indeed lie near the $D \bar D^*$ and $D^* \!
\bar D^*$ thresholds, respectively\footnote{Although we identify the
$Z_c(3900)$ as a $\de$-$\bde$ state, its nature is still fiercely
debated in the literature (among many references, note
Ref.~\cite{Pilloni:2016obd} for its discussion in amplitude analyses
and Ref.~\cite{Ikeda:2016zwx} for a recent lattice simulation).}.
Nevertheless, $Z_c(3900)$ and $Z_c(4020)$  were found in
Ref.~\cite{Giron:2019cfc} to serve naturally as the $I \! = \!  1$,
$J^{PC} \! = \! 1^{+-}$ members of the ground-state $\Sigma^+_g(1S)$
multiplet of the dynamical diquark model, and so in this work we
interpret the two $Z_b$ states analogously.  Furthermore, both
$Z_b(10610)$ and $Z_b(10650)$, like the $Z_c$ states, have been
observed to decay to both closed~\cite{Belle:2011aa} [$\Upsilon(nS)$,
$h_b(nP)$] and open~\cite{Garmash:2015rfd} [$B^{(*)}\bar{B}^{*}$]
heavy-flavor states.  However, the $Z_b$ and $Z_c$ states differ in
one important regard: The observed charmonium decays of $Z_c(3900)$
to date all have total charm-quark spin $s_{c\bar c} \! = \! 1$
({\it i.e.}, that of $J/\psi$), while those of $Z_c(4020)$ have
$s_{c\bar c} \! = \!  0$ ({\it i.e.}, that of {$h_c$), and obtaining
this idealized mixing in a natural way is one of the central results
of Ref.~\cite{Giron:2019cfc}.  However, a glance at
Table~\ref{tab:bbqqExpt} shows that the $Z_b$ system is rather
different: {\em Both\/} $Z_b$ states decay to states with $s_{b\bar b}
\! = \! 0$ and 1 ($\Upsilon$ and $h_b$) with fairly comparable
branching ratios.

The current experimental situation for the hidden-bottom sector also
differs from the hidden-charm sector in one obvious respect: In the
latter, the most obvious and best-studied state is the neutral
$1^{++}$ $X(3872)$.  However, the hidden-bottom analogue $X_b$ ($I^G
\! = \!  0^+$, $J^{PC} \! = \! 1^{++}$) has not yet been observed,
despite a number of searches~\cite{He:2014sqj,Chatrchyan:2013mea,
Aad:2014ama}.  Partly, this absence reflects the relative difficulty
of probing the hidden-bottom sector with limited energy ({\it e.g.},
at the original Belle Experiment operating at a center-of-momentum
energy equal to the $\Upsilon(4S)$ mass~\cite{He:2014sqj}) or limited
only to the decay channel $\Upsilon(1S) \pi^+ \pi^-$ (at the
LHC~\cite{Chatrchyan:2013mea, Aad:2014ama}), which has opposite
$G$-parity to that expected for $X_b$.  It would be truly surprising
in both molecular models and diquark models (as well as in
coupled-channel and QCD sum-rule approaches) were the $X_b$ state to
fail to exist; as a result, a great deal of theoretical effort has
been invested in studying the conjectured
$X_b$~\cite{Tornqvist:1991prl,Tornqvist:1993ng,Tornqvist:2004qy,
Swanson:2006st,Hou:2006it,Ebert:2008se,Ali:2009pi,Guo:2013sya,
Karliner:2013dqa,Chatrchyan:2013mea,Guo:2014sca,Li:2014uia,He:2014sqj,
Aad:2014ama,Karliner:2014lta,Li:2015uwa,Karliner:2015ina,
Patel:2016otd,Wu:2016dws,Zhou:2018hlv,Wang:2019mxn}.  Bounding the
possible range for the $X_b$ mass and determining whether any
hidden-bottom exotics can be even lighter constitute a major goal of
this work.

Table~\ref{tab:bbqqExpt} presents two further observed exotic
candidates, both with $J^{PC} \! = \! 1^{--}$.  The $Y(10750)$ was
recently observed at Belle~\cite{Abdesselam:2019gth}, and has already
been studied as a diquark state~\cite{Ali:2019okl} and within QCD sum
rules~\cite{Wang:2019veq}.  Additionally, $Y(10750)$ and the remaining
exotic candidate $Y(10860)$ have been argued to be conventional
bottomonium states~\cite{Li:2019qsg,Chen:2019uzm}.  One should note,
however, that $Y(10860)$ (like the $Z_b$ states) has both $s_{b\bar b}
\! = \! 0$ and $s_{b\bar b} \! = \! 1$ decay modes (thus violating
heavy-quark spin symmetry in its decays if it is conventional
bottomonium).  In contrast, $Y(10750)$ has thus far been observed to
decay only to $\Upsilon(nS)$.  In addition, $Y(10750)$ lies only
100~MeV above the $Z_b$ states, which would indicate a much smaller
$1P$-$1S$ splitting (assuming they share a related structure) than
between corresponding bottomonium states ({\it e.g.},
$m_{\chi_{bJ}}(1P) \! - \!  m_{\Upsilon(1S)} \! > \!  400$~MeV).  We
argue in Sec.~\ref{sec:Results} that $Y(10860)$ is well suited to
being a $\Sigma^+_g(1P)$ excitation of $\Sigma^+_g(1S)$ states like
$Z_b(10610)$ and $Z_b(10610)$, but $Y(10750)$ is not.

\subsection{The $c\bar{c}s\bar{s}$ Sector}
\label{sec:ccss_ExptReview}
\begin{table*}[ht]
  \caption{All candidate hidden-charm/strange states catalogued by the
  Particle Data Group (PDG)~\cite{Tanabashi:2018oca}.  Also included
  is $Y(4626)$~\cite{Jia:2019gfe,Jia:2020epr}.  Both the particle name
  most commonly used in the literature and its label as given in the
  PDG are listed.}
\label{tab:ccssExpt}
\centering
\setlength{\extrarowheight}{1.5ex}
\begin{tabular}{cccccc}
\hline\hline
    Particle
        & PDG label
            & $I^{G} \, J^{PC}$
                & Mass [MeV]
                    & Width [MeV]
                        & Production and decay \\ \hline
    $X(3915)$ & $\chi_{c0}(3915)$
        & $0^+ \, (0 \, \rm{or} \, 2)^{++}$
            & $3918.4\pm 1.9$
                & $20 \pm 5$
                    & $ e^+e^-\to X$; $X \to\left\{ \begin{array}{l}
                    \omega J/\psi\\
                    \gamma \gamma \\
                    \end{array} \right.$\\
    $Y(4140)$ & $\chi_{c1}(4140)$
        & $0^+ \, 1^{++}$
            & $4146.8 \pm 2.4$
                & $22^{+8}_{-7}$
                    &  $\begin{array}{r} B \to K Y \\
		 p\bar{p} \to Y + {\rm anything} \end{array} \bigg\}$;
                      $Y \to \phi J/\psi$ \\
    $Y(4274)$ & $\chi_{c1}(4274)$
        & $0^+ \, 1^{++}$
            & $4274^{+8}_{-6}$
                & $49 \pm 12$
                    & $B \to K Y$; $Y \to \phi J/\psi$ \\
    $X(4350)$ & $X(4350)$
        & $0^+ \, ?^{?+}$
            & $4351 \pm 5$
                & $ 13^{+18}_{-10}$
                    & $\gamma\gamma \to X$; $X\to \phi J/\psi$ \\
    $X(4500)$ & $\chi_{c0}(4500)$
        & $0^+ \, 0^{++}$
            & $4506^{+16}_{-19}$
                & $ 92 \pm 29$
                    & $p\bar{p} \to X$; $X\to \phi J/\psi$ \\
    $Y(4626)$ & $\psi(4626)$
        & $0^- \, 1^{--}$
            & $4624 \pm 5$
                & $49 \pm 13$
                    & $e^+e^- \to \gamma Y$; $Y \to D_s^+
			D^{\vphantom{+}}_{s1}(2536)^- , \ D_s^+
			D^*_{s2}(2573)^-$ \\ 
    $X(4700)$ & $\chi_{c0}(4700)$
        & $0^+ \, 0^{++}$
            & $4704^{+17}_{-26}$
                & $ 120 \pm 50$
                    & $p\bar{p} \to X$; $X\to \phi J/\psi$ \\
\hline\hline
\end{tabular}
\end{table*}

The most likely hidden-charm/strange ($c\bar c s\bar s$) exotic
candidates are listed in Table~\ref{tab:ccssExpt}.  Almost all have
been seen exclusively in the decay channel $\phi J/\psi$, which
indicates that each either has a valence $c\bar c s\bar s$ quark
content or is a pure $c\bar c$ state decaying through an
Okubo-Zweig-Iizuka (OZI)-suppressed channel.  Similar statements apply
to the newly observed $Y(4626)$~\cite{Jia:2019gfe,Jia:2020epr}, which
has been observed to decay thus far only to channels of open charm and
strangeness.

$X(3915)$ has been included in Table~\ref{tab:ccssExpt} as the
lightest $c\bar c s\bar s$ candidate despite having no observed decays
to states of hidden or open strangeness.\footnote{$X(3915)$ lies below
both the $\phi J/\psi$ and $D_s \bar D_s$ thresholds.  The mode $\eta
\eta_c$ is possible, but here only an upper bound is
known~\cite{Vinokurova:2015txd}.}  Upon its discovery, $X(3915)$ was
immediately assigned by the Particle Data Group (PDG) as the first
radial excitation $\chi_{c0}(2P)$ of the conventional charmonium state
$\chi_{c0}(1P)$.  However, this identification was found to be
problematic for several reasons~\cite{Guo:2012tv,Wang:2014voa,
Olsen:2014maa,Olsen:2018ikz}: First, the mass splitting between
$\chi_{c2}(2P)$ and $X(3915)$ (only about
10~MeV~\cite{Tanabashi:2018oca}) is smaller than the
$\chi_{c2}(2P)$-$\chi_{c0}(2P)$ splitting expected from quark
potential models; furthermore, one would expect $\chi_{c0}(2P)$ (or a
$c\bar c q\bar q$ exotic) to decay prominently into $D \bar{D}$, but
the dominant observed $X(3915)$ decay channel is actually the
OZI-suppressed mode $\omega J/\psi$.  These features led
Ref.~\cite{Lebed:2016yvr} to suppose that $X(3915)$ is actually a
$c\bar{c}s\bar{s}$ state, its $\omega J/\psi$ decay possibly
proceeding by means of a small $s\bar s$ component in $\omega$.
Indeed, the subsequent Belle discovery of
$\chi_{c0}(3860)$~\cite{Chilikin:2017evr} as a candidate with the
expected properties of the missing $\chi_{c0}(2P)$ sharpens the case
for arguing that $X(3915)$ is exotic~\cite{Olsen:2019lcx}.

$S$-wave hidden-charm/strange exotics have been discussed by multiple
authors~\cite{Stancu:2009ka,Wang:2015pea,Li:2015iga,Chen:2016oma,
Wang:2016gxp,Torres:2016oyz,Wang:2016tzr,Liu:2016onn,Maiani:2016wlq,
Wang:2016ujn,Lu:2016cwr,Wu:2016gas,Wang:2016dcb,Agaev:2017foq,
Turkan:2017pil,Chen:2017dpy,Anwar:2018sol,Wang:2018qpe,Wang:2018hwh,
Albuquerque:2018jkn,Yang:2019dxd,Agaev:2020zad}, using methods as
varied as ordinary (tetra)quark models, diquark models,
molecular/rescattering models, and QCD sum rules (as well as
combinations of these).  Following on the observation of the
negative-parity $Y(4626)$, $P$-wave $c\bar c s\bar s$ states have also
recently been considered~\cite{Deng:2019dbg, Zhang:2020gtx}.

The precise nature of the two $1^{++}$ states $Y(4140)$ and $Y(4274)$
in Table~\ref{tab:ccssExpt} is particularly interesting.  On one hand,
they both appear in the mass range predicted for the conventional
$1^{++}$ charmonium state $\chi_{c1}(3P)$.  One might naively think
that since the (missing) $\chi_{c1}(2P)$ state is expected to be quite
wide ($> \! 100$~MeV), its radial excitation $\chi_{c1}(3P)$ should be
even wider.  However, it has been known for some time that the more
complicated wave-function nodal structure of $\chi_{c1}(3P)$ actually
suppresses its width~\cite{Barnes:2005pb} to the same order of
magnitude as that of both $Y(4140)$ and $Y(4274)$.  So then which one,
if either, is the $\chi_{c1}(3P)$?  Studies in which the
$Y(4140)$-$Y(4274)$ sector is described in terms of conventional
charmonium appear in Ref.~\cite{Ortega:2016hde,Chen:2016iua,
Bakker:2019ynk,Hao:2019fjg,Chaturvedi:2019usm,Ferretti:2020civ}.
Moreover, as seen in these papers and in Refs.~\cite{Wang:2014voa,
Liu:2016onn}, no true consensus has emerged on the assignment of
either one.  Additionally, in the simplest diquark models such as the
one used in this work, the ground-state $\Sigma^+_g(1S)$ multiplet
contains only one $1^{++}$ $c\bar c s\bar s$ state [see
Eqs.~(\ref{eq:Swavediquark})].  In this paper, we show that the most
natural assignment identifies $Y(4140)$ as the unique $J^{PC}=1^{++}$
$c\bar c s\bar s$ state and $Y(4274)$ as $\chi_{c1}(3P)$.

\section{Mass Hamiltonian}
\label{sec:MassHamOp}

In the most minimal model variant associated with the dynamical
diquark picture, $b\bar{b} q\bar q^\prime$ exotics ($q,q^\prime \in \{
u, d \}$) connected by a color flux tube in its ground state (the $1S$
multiplet of the $\Sigma^+_g$ BO potential) can be described using a
very simple 3-parameter Hamiltonian:
\begin{eqnarray}
H&=&M_0 + \Delta M_\kqQb +\Delta M_{V_0},\nonumber\\
&=&M_0 + 2\kqQb\left(\mathbf{s}_q \! \cdot\mathbf{s}_b+
\mathbf{s}_{\bar{q}'} \! \cdot\mathbf{s}_{\bar{b}}\right)+V_0
\left(\bm{\tau}_q \! \cdot \! \bm{\tau}_{\bar{q}'}\right)
\left(\bm{\sigma}_q \! \cdot \! \bm{\sigma}_{\bar{q}'}\right) \, .
\nonumber\\
\label{eq:bbqqFullHam}
\end{eqnarray}
Here, $M_0$ is the common $\Sigma^+_g(1S)$ multiplet mass, which
depends only upon the chosen diquark [$\de \! \equiv \! (bq)_{\bf
\bar 3}$ or $\bde \! \equiv \! (\bar b \bar q)_{\bf 3}$] mass and a
central potential $V(r)$ computed numerically on the lattice from pure
glue configurations that connect {\bf 3} and $\bar {\bf 3}$ sources},
as done in Ref.~\cite{Giron:2019bcs}. $M_0$ is the lowest eigenvalue
of the Schr{\"o}dinger equation using the $\Sigma^+_g$ BO potential
$V_{\Sigma^+_g} (r)$; higher eigenvalues have also been computed for
this potential [{\it e.g.}, for $\Sigma^+_g(1P)$, $\Sigma^+_g(2S)$,
{\it etc.}], as well as eigenvalues for lattice-computed excited-glue
configurations ({\it e.g.}, for BO potentials $\Pi^+_u$, $\Sigma^-_u$,
{\it etc.}).

The second term in Eq.~(\ref{eq:bbqqFullHam}) represents the spin-spin
interaction within diquarks, assumed to couple only $q \!
\leftrightarrow \! b$ and $\bar{q}' \! \leftrightarrow \! \bar{b}$, and
$\kqQb$ indicates the strength of this interaction.  These couplings
are singled out as having greater physical effect upon the nature of
the state by assuming that $\de$, $\bde$ are at least somewhat
separated quasiparticles within the full exotic state, so that their
internal spin couplings are expected to be stronger than the ones
between $\de$ and $\bde$.  This ansatz originates with
Ref.~\cite{Maiani:2014aja}, and is incorporated into the motivation
behind the dynamical diquark picture, as described in the Introduction
and discussed in further detail in Refs.~\cite{Brodsky:2014xia,
Giron:2019cfc}.

The final term in Eq.~(\ref{eq:bbqqFullHam}) is an
isospin-spin-dependent interaction between the light-quark spins,
where $V_0$ is the strength of the coupling.  The exotic candidates,
appearing in distinct $I \! = \! 0$ and $I \! = \! 1$ multiplets,
undisputedly exhibit nontrivial isospin dependence, thus requiring a
term such as this to be included in the Hamiltonian.  Its precise form
as given in Eq.~(\ref{eq:bbqqFullHam}) is of course motivated by that
of chiral pion exchanges in hadronic physics, and one plausible
interpretation of this operator~\cite{Giron:2019cfc} is to represent
the effect of exchanging a Goldstone-boson-like mode across the flux
tube connecting the light quarks $q$ and $\bar q^\prime$ in $\de$ and
$\bde$, respectively.  Nevertheless, one could argue for alternate
forms that still carry isospin dependence.  For example,
Refs.~\cite{Giron:2019cfc,Giron:2020fvd} consider the possibility that
the final operator in Eq.~(\ref{eq:bbqqFullHam}) couples not to
light-quark spins ${\bf s}_{q,{\bar q}}$, but to the full diquark
spins ${\bf s}_{\de,\bde}$, which would be an appropriate scheme were
the diquarks truly pointlike.  However, as seen in
Refs.~\cite{Giron:2019cfc,Giron:2020fvd} for the hidden-charm sector,
this alternate formulation leads to results inconsistent with
experiment, such as degeneracy between $X(3872)$ and its (unobserved)
$I \! = \! 1$ partners.

The form of Eq.~(\ref{eq:bbqqFullHam}) has been presented for use in
the $b\bar{b} q\bar q^\prime$ $\Sigma^+_g(1S)$ sector, but as
indicated above, it was originally used for $c\bar{c} q\bar
q^\prime$~\cite{Giron:2019cfc}.  It can be generalized to $B_c$-like
exotics $b\bar{c} q\bar q^\prime$ by using the reduced mass obtained
from unequal $m_\de$ and $m_\bde$ in the Schr{\"o}dinger equation and
introducing unequal $\kappa_{qb}$, $\kappa_{qc}$ coefficients into the
relevant Hamiltonian terms.  Equation~(\ref{eq:bbqqFullHam}) has also
been generalized to the $\Sigma^+_g(1P)$ sector~\cite{Giron:2020fvd}
by the addition of spin-orbit and (isospin-dependent) tensor
couplings.

The ground-state [$\Sigma^+_g(1S)$] hidden-charm/strange ($c\bar c
s\bar s$) exotics can be described using an even simpler Hamiltonian,
since the states lack isospin dependence:
\begin{eqnarray}
H&=&M_0 + \Delta M_\kqQcs,\nonumber\\
&=&M_0 + 2\kqQcs\left(\mathbf{s}_s \! \cdot \mathbf{s}_c+
\mathbf{s}_{\bar{s}} \! \cdot \mathbf{s}_{\bar{c}}\right),
\label{eq:ccssFullHam}
\end{eqnarray}
where $M_0$ and $\kqQcs$ are defined analogously to the parameters
above.  This Hamiltonian actually first appeared in
Ref.~\cite{Lebed:2016yvr}, and also included orbital and spin-orbit
terms to allow comparison between $S$- and $P$-wave states; in the
current model, the $S$-$P$ splitting (as well as the $2S$-$1S$
splitting) can be computed directly using the techniques of
Ref.~\cite{Giron:2019bcs}, as seen in Sec.~\ref{sec:Results}.
Moreover, subsequent experimental findings that confirm the existence
and $J^{PC}$ quantum numbers of relevant states, as well as the
discovery of $X(4500)$ and $X(4700)$ (Table~\ref{tab:ccssExpt}) and
their assignment to the $2S$ multiplet in a diquark
model~\cite{Maiani:2016wlq}, make a fresh analysis of the $c\bar c
s\bar s$ sector quite relevant.

\section{Mass Formula}
\label{sec:MassExp}

The fully general notation for all states in the dynamical diquark
model appears in Ref.~\cite{Lebed:2017min}.  Since the current work
focuses solely on states in the lowest BO potential $\Sigma^+_g$, and
most often those in its lowest multiplet $1S$, we can reduce to a much
more compact notation.  For diquark-antidiquark ($\de$-$\bde$) states
of good total $J^{PC}$ in the $S$-wave band ({\it i.e.}, zero orbital
angular momentum), the defining notation is:
\begin{eqnarray}
J^{PC} = 0^{++}: & \ & X_0 = \left| 0_\de , 0_\bde \right>_0 \, , \ \
X_0^\prime = \left| 1_\de , 1_\bde \right>_0 \, , \nonumber \\
J^{PC} = 1^{++}: & \ & X_1 = \frac{1}{\sqrt 2} \left( \left| 1_\de ,
0_\bde \right>_1 \! + \left| 0_\de , 1_\bde \right>_1 \right) \, ,
\nonumber \\
J^{PC} = 1^{+-}: & \ & Z \  = \frac{1}{\sqrt 2} \left( \left| 1_\de ,
0_\bde \right>_1 \! - \left| 0_\de , 1_\bde \right>_1 \right) \, ,
\nonumber \\
& \ & Z^\prime \, = \left| 1_\de , 1_\bde \right>_1 \, ,
\nonumber \\
J^{PC} = 2^{++}: & \ & X_2 = \left| 1_\de , 1_\bde \right>_2 \, ,
\label{eq:Swavediquark}
\end{eqnarray}
where outer subscripts indicate total quark spin $S \! = \! J$ in the
absence of orbital angular momentum.  The same states may be expressed
in any other spin-coupling basis by using angular momentum recoupling
coefficients, specifically $9j$ symbols.  For both the simplest
evaluation of the final operator in Eq.~(\ref{eq:bbqqFullHam}) and for
convenient physical interpretation, the most useful alternate basis is
that of definite heavy-quark (and light-quark) spin eigenvalues,
$(\QQ) \! + \! (\qq)$:
\begin{eqnarray}
\lefteqn{\left< (s_q \, s_{\bar q}) s_\qq , (s_Q \, s_{\bar Q}) s_\QQ
, S \, \right| \left. (s_q \, s_Q) s_\de , (s_{\bar q} \, s_{\bar Q})
s_\bde , S \right> } & & \nonumber \\
& = & \left( [s_\qq] [s_\QQ] [s_\de] [s_\bde] \right)^{1/2}
\left\{ \begin{array}{ccc} s_q & s_{\bar q} & s_\qq \\
s_Q & s_{\bar Q} & s_\QQ \\ s_\de & s_\bde & S \end{array} \! \right\}
\, , \ \ \label{eq:9jTetra}
\end{eqnarray}
where $[s] \! \equiv \! 2s + 1$ denotes the multiplicity of a spin-$s$
state.  Using Eqs.~(\ref{eq:Swavediquark}) and (\ref{eq:9jTetra}), one
then obtains
\begin{eqnarray}
J^{PC} = 0^{++}: & \ & X_0 = \frac{1}{2} \left| 0_\qq , 0_\QQ
\right>_0 + \frac{\sqrt{3}}{2} \left| 1_\qq , 1_\QQ \right>_0 \, ,
\nonumber \\
& & X_0^\prime = \frac{\sqrt{3}}{2} \left| 0_\qq , 0_\QQ
\right>_0 - \frac{1}{2} \left| 1_\qq , 1_\QQ \right>_0 \, , 
\nonumber \\
J^{PC} = 1^{++}: & \ & X_1 = \left| 1_\qq , 1_\QQ \right>_1 \, ,
\nonumber \\
J^{PC} = 1^{+-}: & \ & Z \; = \frac{1}{\sqrt 2} \left( \left| 
1_\qq , 0_\QQ \right>_1 \! - \left| 0_\qq , 1_\QQ \right>_1 \right)
\, , \nonumber \\
& \ & Z^\prime = \frac{1}{\sqrt 2} \left( \left| 1_\qq ,
0_\QQ \right>_1 \! + \left| 0_\qq , 1_\QQ \right>_1 \right) \, ,
\nonumber \\
J^{PC} = 2^{++}: & \ & X_2 = \left| 1_\qq , 1_\QQ \right>_2 \, .
\label{eq:SwaveQQ}
\end{eqnarray}
A similar recoupling can be used to express these states in terms of
equivalent heavy-light meson spins, $(q \bar Q) + (\bar q Q)$.

The pairs of states $X_0, X^\prime_0$, and $Z, Z^\prime$ carry the
same values of $J^{PC}$ and can therefore mix.  One may define the
equivalent heavy-quark spin eigenstates, which are $X_1$, $X_2$, and
\begin{eqnarray}
{\tilde X}_0 & \equiv & \left| 0_\qq , 0_\QQ \right>_0 =
+ \frac{1}{2} X_0 + \frac{\sqrt{3}}{2} X_0^\prime \, , \nonumber \\
{\tilde X}_0^\prime & \equiv & \left| 1_\qq , 1_\QQ \right>_0 =
+ \frac{\sqrt{3}}{2} X_0 - \frac{1}{2} X_0^\prime \, , \nonumber \\
{\tilde Z} & \equiv & \left| 1_\qq , 0_\QQ \right>_1 =
\frac{1}{\sqrt{2}} \left( Z^\prime \! + Z \right) \, , \nonumber \\
{\tilde Z}^\prime & \equiv & \left| 0_\qq , 1_\QQ \right>_1 =
\frac{1}{\sqrt{2}} \left( Z^\prime \! - Z \right) \, .
\label{eq:HQbasis}
\end{eqnarray}
Assuming $q,q^\prime \in \{ u, d \}$, the $\Sigma^+_g(1S)$ multiplet
for either $c\bar c q\bar q^\prime$ or $b\bar b q\bar q^\prime$ then
consists of precisely 12 isomultiplets: an isosinglet and an
isotriplet corresponding to each of the 6 states in
Eqs.~(\ref{eq:Swavediquark}) or (\ref{eq:SwaveQQ}) [or as reorganized
in Eqs.~(\ref{eq:HQbasis})].  The current PDG
nomenclature~\cite{Tanabashi:2018oca} adopted for the $b\bar b q\bar
q^\prime$ states is $X^{(\prime) \, I=0}_J \! \to \!
\chi^{\vphantom\dagger}_{bJ}$, $X^{(\prime) \, I=1}_J \!
\to \! W^{\vphantom\dagger}_{bJ}$, $Z^{(\prime) \, I=0} \! \to
\!  h_b$, $Z^{(\prime) \, I=1} \! \to \! Z_b$. The corresponding
multiplet for $B_c$-like exotics would also contain 12 isomultiplets,
but which are no longer $C$-parity eigenstates.  If the light quarks
are replaced by an $s\bar s$ pair, then only 6 distinct states remain;
in PDG notation, the $c\bar c s\bar s$ states are labeled
$X^{(\prime)}_J \! \to \! \chi^{\vphantom\dagger}_{cJ}$, $Z^{(\prime)}
\! \to \!  h_c$.

\subsection{Bottomoniumlike Exotics}

Using the Hamiltonian of Eq.~(\ref{eq:bbqqFullHam}) and working (for
definiteness) in the heavy-quark spin basis of
Eqs.~(\ref{eq:HQbasis}), one obtains mass matrices for all 12
isomultiplets of the $b\bar b q\bar q^\prime$ $\Sigma^+_g(1S)$
multiplet.  The cases with nonvanishing off-diagonal elements, for
which the entries are arranged in the order $s_{b\bar b}
\! = \! 0,1$, read
\begin{eqnarray}
\tilde{M}_{0^{++}}^{I=0}&=& M_0
\begin{pmatrix}
1 & 0\\ 0 & 1
\end{pmatrix}
-\kqQb
\begin{pmatrix}
0 & \sqrt{3}\\
\sqrt{3} & 2
\end{pmatrix}
-3V_0
\begin{pmatrix}
-3 & 0\\
0 & 1
\end{pmatrix},\nonumber\\
\tilde{M}_{0^{++}}^{I=1}&=&M_0
\begin{pmatrix}
1 & 0\\
0 & 1
\end{pmatrix}
-\kqQb
\begin{pmatrix}
0 & \sqrt{3}\\
\sqrt{3} & 2
\end{pmatrix}
+V_0
\begin{pmatrix}
-3 & 0\\
0 & 1
\end{pmatrix},\nonumber\\
\tilde{M}_{1^{+-}}^{I=0}&=&
M_0
\begin{pmatrix}
1 & 0\\
0 & 1
\end{pmatrix}
+\kqQb
\begin{pmatrix}
0 & 1\\
1 & 0
\end{pmatrix}
-3V_0
\begin{pmatrix}
1 & 0\\
0 & -3
\end{pmatrix},\nonumber\\
\tilde{M}_{1^{+-}}^{I=1}&=&
M_0
\begin{pmatrix}
1 & 0\\
0 & 1
\end{pmatrix}
+\kqQb
\begin{pmatrix}
0 & 1\\
1 & 0
\end{pmatrix}
+V_0
\begin{pmatrix}
1 & 0\\
0 & -3
\end{pmatrix} .
\end{eqnarray}
Diagonalizing these expressions, and appending the expressions for the
other states (whose mass matrices are already diagonal), one obtains
the mass eigenvalues for all 12 isomultiplets of the $b
\bar b q \bar q^\prime$ $\Sigma^+_g(1S)$ multiplet:
\begin{eqnarray}
M_{0^{++}}^{I=0}&=&\left(M_0-\kqQb +3V_0\right)
\begin{pmatrix}
1 & 0\\
0 & 1
\end{pmatrix}
+2V_1
\begin{pmatrix}
-1 & 0\\
 0 & 1
\end{pmatrix},\nonumber\\
M_{0^{++}}^{I=1}&=&\left(M_0-\kqQb -V_0\right)
\begin{pmatrix}
1 & 0\\
0 & 1
\end{pmatrix}
+2V_2
\begin{pmatrix}
-1 & 0\\
 0 & 1
\end{pmatrix},\nonumber\\
M_{1^{+-}}^{I=0}&=&\left(M_0+3V_0\right)
\begin{pmatrix}
1 & 0\\
0 & 1
\end{pmatrix}
+V_3
\begin{pmatrix}
-1 & 0\\
 0 & 1
\end{pmatrix},\nonumber\\
M_{1^{+-}}^{I=1}&=&\left(M_0-V_0\right)
\begin{pmatrix}
1 & 0\\
0 & 1
\end{pmatrix}
+V_4
\begin{pmatrix}
-1 & 0\\
 0 & 1
\end{pmatrix},\nonumber\\
M_{1^{++}}^{I=0}&=&M_0-\kqQb -3V_0 \, ,\nonumber\\
M_{1^{++}}^{I=1}&=&M_0-\kqQb +V_0 \, ,\nonumber\\
M_{2^{++}}^{I=0}&=&M_0+\kqQb -3V_0 \, ,\nonumber\\
M_{2^{++}}^{I=1}&=&M_0+\kqQb +V_0 \, ,
\label{eq:AllMasses}
\end{eqnarray}
where we abbreviate
\begin{eqnarray}
V_1 & \equiv & \sqrt{\kqQb^2 +3\kqQb V_0 +9V_0^2} \, , \nonumber \\
V_2 & \equiv & \sqrt{\kqQb^2 -\kqQb V_0 +V_0^2} \, , \nonumber \\
V_3 & \equiv & \sqrt{\kqQb^2 +36V_0^2} \, , \nonumber \\
V_4 & \equiv & \sqrt{\kqQb^2 +4V_0^2} \, . \label{eq:Vparams}
\end{eqnarray}
The pairs of states in Eqs.~(\ref{eq:AllMasses}) degenerate in
$J^{PC}$ are arranged in order of increasing mass.

To obtain the mixing angles, one must first derive the corresponding
normalized eigenvectors for the 4 mixed pairs of states with
$J^{PC}=0^{++},\;1^{+-}$.  Further denoting
\begin{equation}
\epsilon_{qQ} \equiv {\rm sgn} (\kappa_{qQ}) \, ,
\end{equation}
the normalized eigenvectors collected into columns of unitary matrices
$R$ read
\begin{eqnarray}
R^{I=0}_{0^{++}} & = & \frac{1}{2\sqrt{V_1}}
\nonumber\\
\lefteqn{\hspace{-3.1em} \times \begin{pmatrix}
\ \ \ \ \sqrt{2V_1-(\kqQb +6V_0)} &
\sqb \sqrt{2V_1+(\kqQb +6V_0)}\\
\sqb \sqrt{2V_1+(\kqQb +6V_0)}&
\ \ -\sqrt{2V_1-(\kqQb +6V_0)}
\end{pmatrix},}\nonumber\\
R^{I=1}_{0^{++}} & = & \frac{1}{2\sqrt{V_2}}
\nonumber\\
\lefteqn{\hspace{-3.1em} \times \begin{pmatrix}
\ \ \ \ \sqrt{2V_2-(\kqQb -2V_0)}&
\sqb \sqrt{2V_2+(\kqQb -2V_0)}\\
\sqb \sqrt{2V_2+(\kqQb -2V_0)}&
\ \ -\sqrt{2V_2-(\kqQb -2V_0)}
\end{pmatrix},}\nonumber\\
R^{I=0}_{1^{+-}} &=&\frac{1}{\sqrt{2V_3}}
\begin{pmatrix}
\sqb \sqrt{V_3+6V_0} &
\ \ \ \ \sqrt{V_3-6V_0}\\
\ \ -\sqrt{V_3-6V_0} &
\sqb \sqrt{V_3+6V_0}
\end{pmatrix},\nonumber\\
R^{I=1}_{1^{+-}} &=&\frac{1}{\sqrt{2V_4}}
\begin{pmatrix}
\sqb \sqrt{V_4-2V_0} &
\ \ \ \ \sqrt{V_4+2V_0}\\
\ \ -\sqrt{V_4+2V_0} &
\sqb \sqrt{V_4-2V_0}
\end{pmatrix}. \; \label{eq:bbqqEigenvectors}
\end{eqnarray}
The probability $P$ of the lighter mass eigenstate in each mixed case
to be measured to have heavy-quark spin eigenvalue $s_{b\bar{b}}
\! = \! 1$, which is simply obtained by squaring the 1,2 element in
each matrix of Eqs.~(\ref{eq:bbqqEigenvectors}), is given by
\begin{eqnarray}
P_{0^{++}, \, s_{b\bar{b}}=1}^{I=0}&=&  \frac{1}{2} +
\frac{\kqQb + 6V_0}{4\sqrt{\kqQb^2 +3\kqQb V_0 + 9V_0^2}}
\, , \nonumber \\
P_{0^{++}, \, s_{b\bar{b}}=1}^{I=1}&=& \frac{1}{2} +
\frac{\kqQb - 2V_0}{4\sqrt{\kqQb^2 -\kqQb V_0 +V_0^2}} \, , \nonumber
\\
P_{1^{+-}, \, s_{b\bar{b}}=1}^{I=0}&=& \frac{1}{2} -
\frac{3V_0}{\sqrt{\kqQb^2 +36V_0^2}} \, , \nonumber\\
P_{1^{+-}, \, s_{b\bar{b}}=1}^{I=1}&=& \frac{1}{2} +
\frac{V_0}{\sqrt{\kqQb^2 +4V_0^2}} \, . \label{eq:Pvalues}
\end{eqnarray}
Assuming that heavy-quark symmetry is unbroken in the decays of these
states, the $P$ values give the relative branching ratios for the
lighter mass eigenstate in each case to decay into a bottomonium state
with $s_{b\bar{b}} \!  = \! 1$ ($\Upsilon$, $\chi_b$) {\it vs.\@}
$s_{b\bar{b}} \! = \!  0$ ($\eta_b$, $h_b$).

\subsection{Hidden-Charm/Strange Exotics}

Using the Hamiltonian of Eq.~(\ref{eq:ccssFullHam}) and working (for
definiteness) in the heavy-quark spin basis of
Eqs.~(\ref{eq:HQbasis}), one obtains mass matrices for all 6 states of
the $c\bar c s\bar s$ $\Sigma^+_g(1S)$ multiplet.  The cases with
nonvanishing off-diagonal elements, for which the entries are arranged
in the order $s_{c\bar c} \! = \! 0,1$, read
\begin{eqnarray}
\tilde{M}_{0^{++}}&=& M_0
\begin{pmatrix}
1 & 0\\
0 & 1
\end{pmatrix}-\kqQcs
\begin{pmatrix}
0 & \sqrt{3}\\
\sqrt{3} & 2
\end{pmatrix},\nonumber\\
\tilde{M}_{1^{+-}}&=&M_0
\begin{pmatrix}
1 & 0\\
0 & 1
\end{pmatrix}+\kqQcs
\begin{pmatrix}
0 & 1\\
1 & 0
\end{pmatrix} .
\end{eqnarray}
Diagonalizing these expressions, and appending the expressions for the
other states (whose mass matrices are already diagonal), one obtains
the mass eigenvalues for all 6 states of the $c\bar c s\bar s$
$\Sigma^+_g(1S)$ multiplet:
\begin{eqnarray}
M_{0^{++}}&=&\left(M_0-\kqQcs \right)
\begin{pmatrix}
1 & 0\\
0 & 1
\end{pmatrix}+2|\kqQcs |
\begin{pmatrix}
-1 & 0\\
0 & 1
\end{pmatrix},\nonumber\\
M_{1^{+-}}&=& M_0
\begin{pmatrix}
1 & 0\\
0 & 1
\end{pmatrix}+|\kqQcs |
\begin{pmatrix}
-1 & 0\\
0 & 1
\end{pmatrix},\nonumber\\
M_{1^{++}}&=&M_0-\kqQcs ,\nonumber\\
M_{2^{++}}&=&M_0+\kqQcs . \label{eq:ccssMasses}
\end{eqnarray}
The pairs of states in Eqs.~(\ref{eq:ccssMasses}) degenerate in
$J^{PC}$ are arranged in order of increasing mass.  Note at this point
we have not constrained the spin-spin coupling $\kqQcs$ to assume a
positive value.

To obtain the mixing angles, one must first derive the corresponding
normalized eigenvectors for the 2 mixed pairs of states with
$J^{PC}=0^{++},\;1^{+-}$.  Collected into columns of unitary matrices
$R$, the eigenvectors read
\begin{eqnarray}
R_{0^{++}} &=&\frac{1}{2}\begin{pmatrix}
\ \ \ \ \sqrt{2 -\ssc} & \ssc \sqrt{2 + \ssc }\\
\ssc \sqrt{2 + \ssc } & \ \ -\sqrt{2 - \ssc }
\end{pmatrix},\nonumber\\
R_{1^{+-}} &=&\frac{1}{\sqrt{2}}
\begin{pmatrix}
\ \ \ssc & 1\\
-1 & \ssc
\end{pmatrix}. \label{eq:ccssEigenvectors}
\end{eqnarray}

The probability $P$ of the lighter mass eigenstate in each mixed case
to be measured to have heavy-quark spin eigenvalue $s_{c\bar{c}}
\! = \! 1$, which is simply obtained by squaring the 1,2 element in
each matrix of Eqs.~(\ref{eq:ccssEigenvectors}), is given by
\begin{eqnarray}
P_{0^{++}, \, s_{c\bar{c}}=1}&=& \frac{1}{2} + \frac{1}{4} \ssc ,
\nonumber\\
P_{1^{+-}, \, s_{c\bar{c}}=1}&=& \frac{1}{2}.
\end{eqnarray}
Assuming that heavy-quark symmetry is unbroken in the decays of these
states, the $P$ values give the relative branching ratios for the
lighter mass eigenstate in each case to decay into a charmonium state
with $s_{c\bar{c}} \!  = \! 1$ ($\psi$, $\chi_c$) {\it vs.\@}
$s_{c\bar{c}} \! = \! 0$ ($\eta_c$, $h_c$).

\section{Analysis and Results}
\label{sec:Results}

\subsection{$c\bar{c}q\bar{q}^\prime$ Exotics Redux}

The masses of the 12 isomultiplets in the $b\bar b q\bar q^\prime$
$\Sigma^+_g(1S)$ multiplet depend upon only 3 Hamiltonian parameters:
$M_0$, $\kqQb$, and $V_0$, as seen in
Eqs.~(\ref{eq:AllMasses})--(\ref{eq:Vparams}).  A similar, but not
identical, analysis of the 12 $c\bar c q\bar q^\prime$
$\Sigma^+_g(1S)$ isomultiplets appears in Ref.~\cite{Giron:2019cfc}
(with, of course, $\kqQb \! \to \! \kappa_{qc}$, and different $M_0$
and $V_0$ numerical values for the $c\bar c q\bar q^\prime$ and $b\bar
b q\bar q^\prime$ systems).  There, the masses of the 3 states
$X(3872)$, $Z_c(3900)$, and $Z_c(4020)$~\cite{Tanabashi:2018oca} are
used as inputs, and the mixing angles of $0^{++}$ and $1^{+-}$ states
are allowed to vary under the reasoning that any additional operators
omitted from the minimal 3-parameter form have small numerical
coefficients and would leave the mass spectrum stable, but could
nevertheless substantially change the precise values of the mixing
angles.  Using the additional phenomenological observation that
$X_1^{I=0}$ [corresponding to $X(3872)$] appears to be the lightest
$c\bar c q\bar q^\prime$ state, Ref.~\cite{Giron:2019cfc} obtained
\begin{equation}
M_0 = 3988.75 \, {\rm MeV} , \ \kappa_{qc} = 17.76 \, {\rm MeV} , \
V_0 = 33.10 \, {\rm MeV} . \label{eq:JHEPnums}
\end{equation}
From these values, Ref.~\cite{Giron:2019cfc} found that $Z_c(3900)$
decays almost exclusively to $J/\psi$ and $Z_c(4020)$ to $h_c$, in
full accord with current observations.

However, one may just as easily adopt the strict 3-parameter form of
Eq.~(\ref{eq:bbqqFullHam}) for the $c\bar c q\bar q^\prime$ sector,
and use the 3 measured mass eigenvalues for $M_{1^{++}}^{I=0}$ and
$M_{1^{+-}}^{I=1}$ in Eqs.~(\ref{eq:AllMasses})--(\ref{eq:Vparams}) to
obtain values for the parameters $M_0$, $\kappa_{qc}$, and $V_0$, as
well as for the mixing parameters $P$ of Eqs.~(\ref{eq:Pvalues}).  A
double-valued set of equations then arises; one solution gives nearly
identical values to Eq.~(\ref{eq:JHEPnums}):
\begin{equation}
M_0 = 3988.69 \; {\rm MeV} , \ \kappa_{qc} = 17.89 \; {\rm MeV} , \
V_0 = 33.04 \; {\rm MeV} , \label{eq:ccqqFit}
\end{equation}
and the very satisfactory value $P_{s_{c\bar c} = 1}[Z_c(3900)] \! =
\! 0.983$.  The other solution gives rather different values:
\begin{equation}
M_0 = 3964.59 \; {\rm MeV} , \ \kappa_{qc} = 66.07 \; {\rm MeV} , \
V_0 = 8.94 \; {\rm MeV} ,
\end{equation}
and the phenomenologically unacceptably small value $P_{s_{c\bar c} =
1}[Z_c(3900)] \! = \! 0.631$.  One learns from this exercise that the
value of $P$, even if not precisely measured, serves as a decisive
input to the model.

But one also finds, using the fit values of Eqs.~(\ref{eq:ccqqFit}) in
the minimal 3-parameter model, that $X_1^{I=0}$ is no longer the
lightest $c\bar c q\bar q^\prime$ state; $X_0^{I=0}$ (the $0^{++}$
isosinglet) assumes that status, with
\begin{equation}
M_{X_0^{I=0}} = 3851.6 \; {\rm MeV} \, .
\end{equation}
This prediction is remarkable, in that it overlaps with the observed
mass $3862^{+50}_{-35}$~MeV of the conventional charmonium
$\chi_{c0}(2P)$ candidate~\cite{Chilikin:2017evr}, which shares the
same quantum numbers.  The large observed width
$201^{+180}_{-110}$~MeV indicates unimpeded $S$-wave decays into $D
\bar D$ pairs (threshold $\approx$~3740~MeV) for either
$\chi_{c0}(2P)$ or $X_0^{I=0}$, and indeed, the observed
$\chi_{c0}(3860)$ could be a mixture of the two.

\subsection{Bottomoniumlike Exotics}
\label{sec:bbqqFits}

\begin{table*}[t]
\centering
\caption{Predictions for the 12 isomultiplet masses (in MeV) of the
$\Sigma^+_g(1S)$ $b\bar b q\bar q^\prime$ multiplet, using the
Hamiltonian of Eq.~(\ref{eq:bbqqFullHam}) as evaluated using
Eqs.~(\ref{p_1_pm}), (\ref{eq:MassesWithP}), and
(\ref{eq:AbbrevsWithP}).  Boldface indicates the measured $Z_b$ mass
inputs.}
\setlength{\extrarowheight}{1.0ex}
\begin{tabular}{c | c c | c c |c c | c c | c c | c c}
\hline\hline
& \multicolumn{4}{c|}{$P = P_{\bar Z , \, s_{b\bar{b}}=1}^{I=1}=1/4$}
&\multicolumn{4}{c|}{$P = P_{\bar Z , \, s_{b\bar{b}}=1}^{I=1}=1/2$}&
\multicolumn{4}{c}{$P = P_{\bar Z , \, s_{b\bar{b}}=1}^{I=1}=3/4$}\\
$J^{PC}$ & \multicolumn{2}{c}{$I=0$} & \multicolumn{2}{c|}{$I=1$}&
\multicolumn{2}{c}{$I=0$} & \multicolumn{2}{c|}{$I=1$} &
\multicolumn{2}{c}{$I=0$} & \multicolumn{2}{c}{$I=1$} \\
\hline
$0^{++}$& $10551.1$ & $10624.4$ & $10564.6$ & $10655.9$ & $10562.2$ &
$10652.2$ & $10562.2$ & $10652.2$ & $10569.7$ & $10695.7$ & $10575.4$
& $10644.9$\\
$1^{++}$& $10621.5$ & $ $ & $10599.0$ & $ $ & $10607.2$ & $ $ &
$10607.2$ & $ $ &$10598.9$ & $ $ & $10621.4$ & $ $\\
$1^{+-}$& $10568.3$ & $10646.2$ & $\mathbf{10607.2}$ &
$\mathbf{10652.2}$ & $10607.2$ & $10652.2$ & $\mathbf{10607.2}$ &
$\mathbf{10652.2}$& $10613.2$ & $10691.1$ & $\mathbf{10607.2}$ &
$\mathbf{10652.2}$\\
$2^{++}$& $10660.5$ & $ $ & $10638.0$ & $ $ & $10652.2$ & $ $ &
$10652.2$ & $ $  & $10637.9$ & $ $ & $10660.4$ & $ $\\
\hline\hline
\end{tabular}
\label{tab:bbqq_mass_predictions}
\end{table*}
Table~\ref{tab:bbqqExpt} shows that only 2 out of 12 $b\bar b q\bar
q^\prime$ candidates in the positive-parity $\Sigma^+_g(1S)$ multiplet
have been observed to date, both with $(I^G)\;J^{PC} \! = \!
(1^+)\;1^{+-}$: $Z_b(10610)$ and $Z_b(10650)$.  Two known masses for a
model with 3 Hamiltonian parameters hardly seems sufficient input to
draw many conclusions, but the results of the previous subsection
indicate that using the $s_{b\bar b} \! = \! 1$ content $P_{1^{+-}, \,
s_{b\bar{b}}=1}^{I=1}$ of $Z_b(10610)$ can be helpful.  Indeed, we
define
\begin{eqnarray}
\Mb0 &\equiv & M_0 - V_0 \nonumber\\
&=& \frac{1}{2}\left[m_{Z_b(10650)}+m_{Z_b(10610)}\right]
=10629.7\;\rm{MeV} , \;\;\;\; \nonumber \\
V_4 &\equiv & \sqrt{\kqQb^2 +4V_0^2} \nonumber\\
&=& \frac{1}{2}\left[m_{Z_b(10650)}-m_{Z_b(10610)}\right]
=22.5\;\rm{MeV} , \nonumber \\
P &\equiv & P_{1^{+-}, \, s_{b\bar{b}}=1}^{I=1}\nonumber\\
&=&  \frac{1}{2} + \frac{V_0}{\sqrt{\kqQb^2 +4V_0^2}}
=\frac{1}{2}+\frac{V_0}{V_4} \, , \label{p_1_pm}
\end{eqnarray}
where the definitions of $V_4$ and $P_{1^{+-} , \,
s_{b\bar{b}}=1}^{I=1}$ are the same as in Eqs.~(\ref{eq:Vparams}) and
(\ref{eq:Pvalues}), respectively, and for definiteness our numerical
analysis uses the mass of the charged $Z_b(10650)$.  Using these
definitions, one may express the original parameters in
Eq.~(\ref{eq:bbqqFullHam}) as
\begin{eqnarray}
M_0 & = & \Mb0 + V_4\left(P-\frac{1}{2}\right) \, , \nonumber \\
|\kqQb| &=& 2V_4\sqrt{P\left(1-P\right)} \, , \nonumber \\
V_0 &=& V_4\left(P-\frac{1}{2}\right) \, . \label{v0_r_p}
\end{eqnarray}
Given a particular numerical value for $P$, the only remaining
ambiguity in predicting the entire $\Sigma^+_g(1S)$ mass spectrum is
the sign of $\kqQb$.  With reference to Eq.~(\ref{eq:bbqqFullHam}),
$\kqQb \! > \! 0$ indicates a scenario in which the spin-singlet
diquark $\de \! \equiv (qb)$ is lighter than the spin-triplet, which
is the expectation of virtually every model.  Thus making the mild
assumption that $\kqQb \! > \! 0$, the formulas of
Eqs.~(\ref{eq:AllMasses})--(\ref{eq:Vparams}) for the mass eigenstates
(indicated henceforth by overlines, with primes for the heavier of
states that are degenerate in $J^{PC}$) then read
\begin{eqnarray}
M_{\bar{X}_0}^{I=0}&=& \Mb0 -V_4\left[2\sqrt{P\left(1-P\right)}
-2\left(2P-1\right) \! + C_1(P) \right],\nonumber\\
M_{\bar{X}_0'}^{I=0}&=&\Mb0 -V_4\left[2\sqrt{P\left(1-P\right)}
-2\left(2P-1\right) \! - C_1(P) \right],\nonumber\\
M_{\bar{X}_0}^{I=1}&=&\Mb0 - V_4\left[2\sqrt{P\left(1-P\right)}
+ C_2(P) \right],\nonumber\\
M_{\bar{X}_0'}^{I=1}&=& \Mb0 - V_4\left[2\sqrt{P\left(1-P\right)}
- C_2(P) \right],\nonumber\\
M_{\bar{Z}}^{I=0}&=&\Mb0 + V_4\left[2\left(2P-1\right)
-\sqrt{9-32P\left(1-P\right)}\right],\nonumber\\
M_{\bar{Z}'}^{I=0}&=&\Mb0 + V_4\left[2\left(2P-1\right)
+\sqrt{9-32P\left(1-P\right)}\right],\nonumber\\
M_{\bar{Z}}^{I=1}&=& \Mb0 -V_4 = M_{Z_b(10610)}, \nonumber\\
M_{\bar{Z}'}^{I=1}&=& \Mb0 +V_4 = M_{Z_b(10650)},\nonumber\\
M_{X_1}^{I=0}&=& \Mb0 - V_4\left[2\sqrt{P\left(1-P\right)}
+\left(2P-1\right)\right],\nonumber\\
M_{X_1}^{I=1}&=& \Mb0 - V_4\left[2\sqrt{P\left(1-P\right)}
-\left(2P-1\right)\right],\nonumber\\
M_{X_2}^{I=0}&=& \Mb0 + V_4\left[2\sqrt{P\left(1-P\right)}
-\left(2P-1\right)\right],\nonumber\\
M_{X_2}^{I=1}&=& \Mb0 + V_4\left[2\sqrt{P\left(1-P\right)}
+\left(2P-1\right)\right], \label{eq:MassesWithP}
\end{eqnarray}
where we abbreviate
\begin{eqnarray}
C_1(P) & \equiv & \sqrt{9-20P\left(1-P\right)+12\left(2P-1\right)
\sqrt{P\left(1-P\right)}} \, , \nonumber \\
C_2(P) & \equiv & \sqrt{1+12P\left(1-P\right)-4\left(2P-1\right)
\sqrt{P\left(1-P\right)}} \, . \nonumber \\ \label{eq:AbbrevsWithP}
\end{eqnarray}
The expressions in Eqs.~(\ref{eq:Pvalues}) for the heavy-quark spin
content of the remaining mixed states then assume the forms
\begin{eqnarray}
P_{\bar X_0 , \, s_{b\bar{b}}=1}^{I=0}&=&\frac{1}{2}
\left[1+\frac{2\sqrt{P\left(1-P\right)}+3\left(2P-1\right)}
{C_1(P)}
\right],\nonumber\\
P_{\bar X_0 , \, s_{b\bar{b}}=1}^{I=1}&=&\frac{1}{2}
\left[1+\frac{2\sqrt{P\left(1-P\right)}-\left(2P-1\right)}
{C_2(P)}
\right],\nonumber\\
P_{\bar Z , \, s_{b\bar{b}}=1}^{I=0}&=&\frac{1}{2}
\left[1-\frac{3\left(2P-1\right)}
{\sqrt{9-32P\left(1-P\right)}}\right] , \label{eq:AllPValues}
\end{eqnarray}
and all observables for the entire $\Sigma^+_g(1S)$ multiplet are now
expressed as functions of the single parameter $P \! \equiv \! P_{\bar
Z , s_{b\bar b = 1}}^{I=0}$, which varies between 0 and 1; the only
numerical inputs are the $Z_b(10610)$ and $Z_b(10650)$ masses.

In fact, sufficient data exists to go even further: An examination of
the exclusive $\Upsilon$- and $h_b$-channel branching ratios in
Table~\ref{tab:bbqqExpt} reveals some interesting effects.  First, the
branching ratios to $\Upsilon(1S)$ are the smallest among bottomonium
decays for both $Z_b(10610)$ and $Z_b(10650)$, and the branching
ratios to $h_b(2S)$ are the largest.  Noting from simple
quark-potential models that $\Upsilon(1S)$ has by far the most
spatially compact bottomonium wave function while $h_b(2P)$ has the
largest of those kinematically allowed in the $Z_b$ decays, one is led
to the qualitative conclusion that the $Z_b$ states are not spatially
compact.  Moreover, $h_b(2P)$ has a complicated wave function with not
only angular but radial nodes, suggesting initial $Z_b$ wave functions
that are similarly nonuniform in their spatial density.  For our
immediate purposes, however, the most interesting feature arises in a
direct comparison of the branching ratios for individual $\Upsilon$
and $h_b$ channels, noting that the phase space factors for exclusive
$Z_b(10610)$ and $Z_b(10650)$ decay modes are almost identical.  With
the possible exception of the $\Upsilon(3S)$, the $\Upsilon$ branching
ratios of the $Z_b(10610)$ appear to be a factor of about 3 times
larger than those of the $Z_b(10650)$, and the $h_b$ branching ratios
of the $Z_b(10610)$ appear to be a factor of about 3 times smaller
than those of the $Z_b(10650)$.  One is therefore led to the natural
estimate $P \! \approx \! 3/4$.

In addition, the last of Eqs.~(\ref{v0_r_p}) shows that the sign of $P
\! - \! \frac 1 2$ directly gives the sign of $V_0$.  Since as mentioned
below its definition in Eq.~(\ref{eq:bbqqFullHam}), the $V_0$ term is
motivated~\cite{Giron:2019cfc} by its similarity in form to the
attractive pion interaction in hadronic physics, the value of $P
\! \approx \! 3/4$ obtained above gives $V_0 \! > \! 0$ and
suggests a similar interaction in the $\Sigma^+_g(1S)$ multiplet for
both charmoniumlike and bottomoniumlike states.  Based upon these
consideration, we expect $1/2 \! < \! P \! < \! 1$.  In fact,
independently of $V_0$, the values of $M_0$ and $\kqQb$ obtained
solely from the $Z_b$ masses are numerically very stable over the
entire range of $P$, and demonstrate this fact by exhibiting results
at $P \! = \! 1/4$ in addition to $P \! = \! 1/2$ and 3/4:
\begin{eqnarray}
P=\frac{1}{4} \rightarrow
M_0   & = & 10624.08\;\rm{MeV},\nonumber\\
\kqQb & = & 19.49\;\rm{MeV},\nonumber\\
V_0   & = & -5.63\;\rm{MeV},\\
P=\frac{1}{2} \rightarrow
M_0   & = & 10629.70\;\rm{MeV},\nonumber\\
\kqQb & = & 22.50\;\rm{MeV},\nonumber\\
V_0   & = & 0.00\;\rm{MeV},\\
P=\frac{3}{4} \rightarrow
M_0   & = & 10635.33\;\rm{MeV},\nonumber\\
\kqQb & = & 19.49\;\rm{MeV},\nonumber\\
V_0   & = & +5.63\;\rm{MeV}. \label{eq:bbParamValues}
\end{eqnarray}
Of special note, the allowed values of $\kqQb$ in this range of $P$
are numerically very close to those obtained in Eq.~(\ref{eq:ccqqFit})
for $\kappa_{qc}$, indicating a common physical origin for both.  In
contrast, the allowed values of $V_0$ are several times smaller in the
hidden-bottom sector, reflecting the expectation that $V_0$ contains a
coefficient scaling inversely with a power of the heavy-quark mass and
therefore being smaller for bottomoniumlike than for charmoniumlike
states.

Inserting the values of parameters determined in
Eqs.~(\ref{eq:bbParamValues}) from these three choices of $P$, we
obtain predictions for masses of all 12 isomultiplets of the $b\bar b
q\bar q^\prime$ $\Sigma^+_g(1S)$ multiplet in
Table~\ref{tab:bbqq_mass_predictions}.  The most notable feature of
these results is the remarkably small numerical variation of
individual state mass predictions over the whole range $1/4 \! \leq \!
P \! \leq \! 3/4$, keeping in mind that the expected sign of $V_0$
and the decay pattern disfavor $P \! < \! 1/2$. 

\begin{figure}
\centering
\includegraphics[width=\columnwidth]{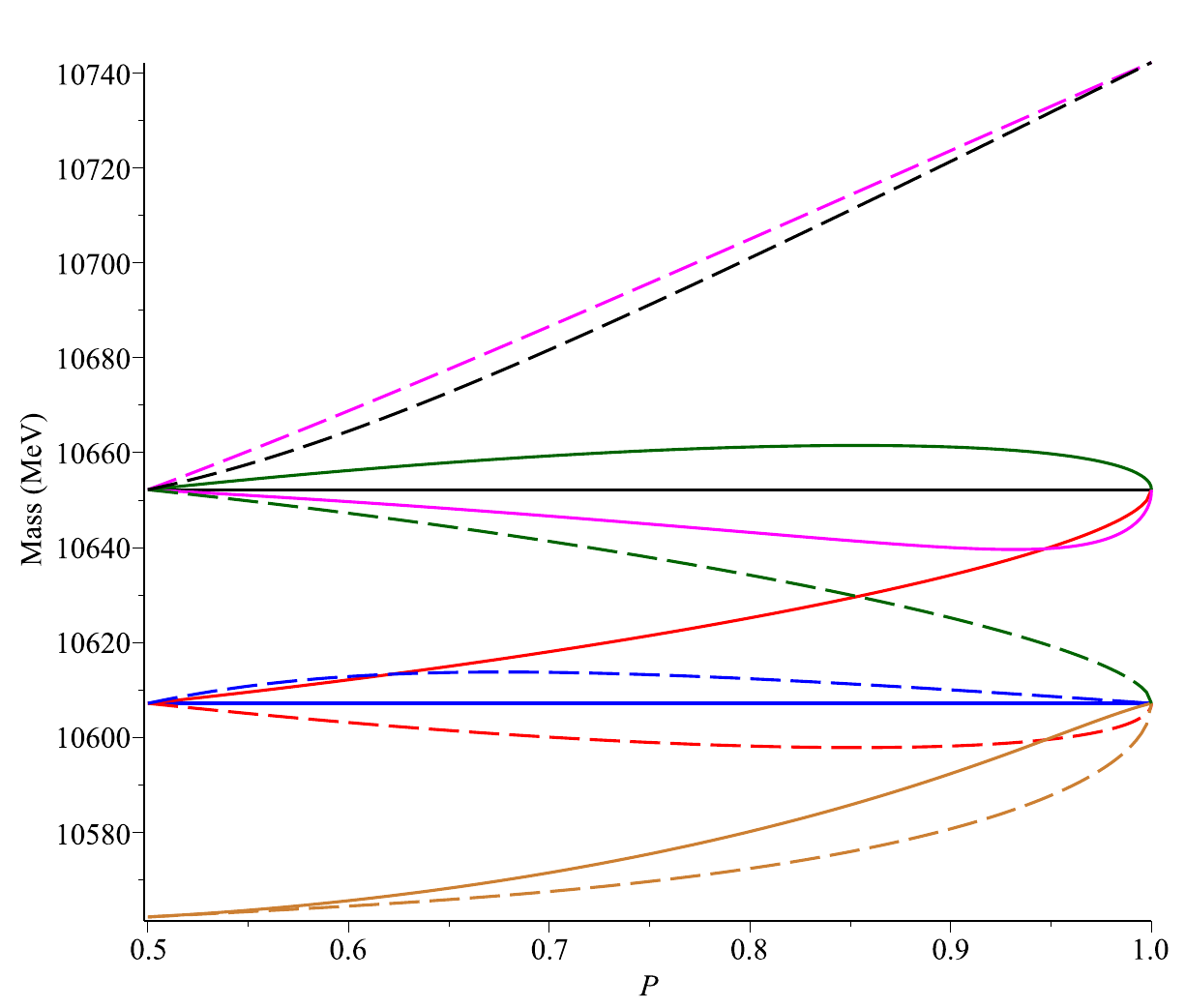}
\caption{(Color online)  Prediction of the 12 isomultiplet masses
(in MeV) of the $\Sigma^+_g(1S)$ multiplet as functions of the
heavy-quark $s_{b\bar b} \! = \! 1$ spin-content parameter $P$ of
$Z_b(10610)$ defined in Eq.~(\ref{p_1_pm}).  Solid (dashed) lines
indicate $I \! = \! 1$ ($I \! = \! 0$) states.  Using the naming
scheme of Eq.~(\ref{eq:Swavediquark}) with isospin superscripts, an
overline for mass eigenstates, and a prime for the heavier of mixed
eigenstates, the levels from top to bottom at $P \! = \! 3/4$ are:
$\bar X^{\prime \, I=0}_0$ (dashed magenta); $\bar Z^{\prime \, I=0}$
(dashed black); $X^{I=1}_2$ (solid green); $\bar Z^{\prime \, I=1}$
(solid black); $\bar X^{\prime \, I=1}_0$ (solid magenta); $X^{I=0}_2$
(dashed green); $X^{I=1}_1$ (solid red); $\bar Z^{I=0}$ (dashed blue);
$\bar Z^{I=1}$ (solid blue); $X^{I=0}_1$ (dashed red); $\bar
X^{I=1}_0$ (solid gold); $\bar X^{I=0}_0$ (dashed gold).
\label{fig:MassSpectrum_P}
}
\end{figure}

Another way to visualize these results is to impose our expectation
that $P \! \ge 1/2$ and consider the entire range $1/2 \!
\leq \! P \! \leq \!  1$.  We then plot the results from combining
Eqs.~(\ref{p_1_pm}), (\ref{eq:MassesWithP}), and
(\ref{eq:AbbrevsWithP}) for all 12 $\Sigma^+_g(1S)$ isomultiplet
masses in Fig.~\ref{fig:MassSpectrum_P}.  The ordering of the states
in this range of $P$ is remarkably stable.  Of particular note: Over
most of the allowed range for $P$, the isosinglet $J^{PC} \! = \!
0^{++}$ state $\bar X_0^{I=0}$ is lightest, and its isotriplet partner
$\bar X_0^{I=1}$ is second lightest.  Both lie above the threshold
($\approx \!  10560$~MeV) of their expected dominant $B\bar B$ decay
channel but not excessively so, suggesting that reasonably narrow
$0^{++}$ $b\bar b q\bar q^\prime$ states will be discovered in future
experiments.  Meanwhile, $\bar X_b \! \equiv \! X_1^{I=0}$, the $b\bar
b q\bar q^\prime$ analogue to the $X(3872)$, only becomes the
second-lightest $b\bar b q\bar q^\prime$ state for values of $P$ very
close to 1 (which is what occurs in the $c\bar c q\bar q^\prime$
system).  More interestingly, $\bar X^{I=0}_1$ lies at most only a few
MeV below the $B\bar B^*$ threshold ($\approx \! 10605$~MeV) over
almost the whole range $1/2 \! \leq \!  P \! \leq \! 1$, and thus
analogously to $X(3872)$ in its relation to $D\bar D^*$, $\bar
X^{I=0}_1$ will need to be analyzed by considering the impact of
$B\bar B^*$ threshold effects.  Explicitly, we predict
\begin{equation}
10598 \ {\rm MeV} \leq m_{X_b} \equiv m_{\bar X^{I=0}_1} \leq 10607 \
{\rm MeV} \, .
\end{equation}

\begin{figure}
\centering
\includegraphics[width=\columnwidth]{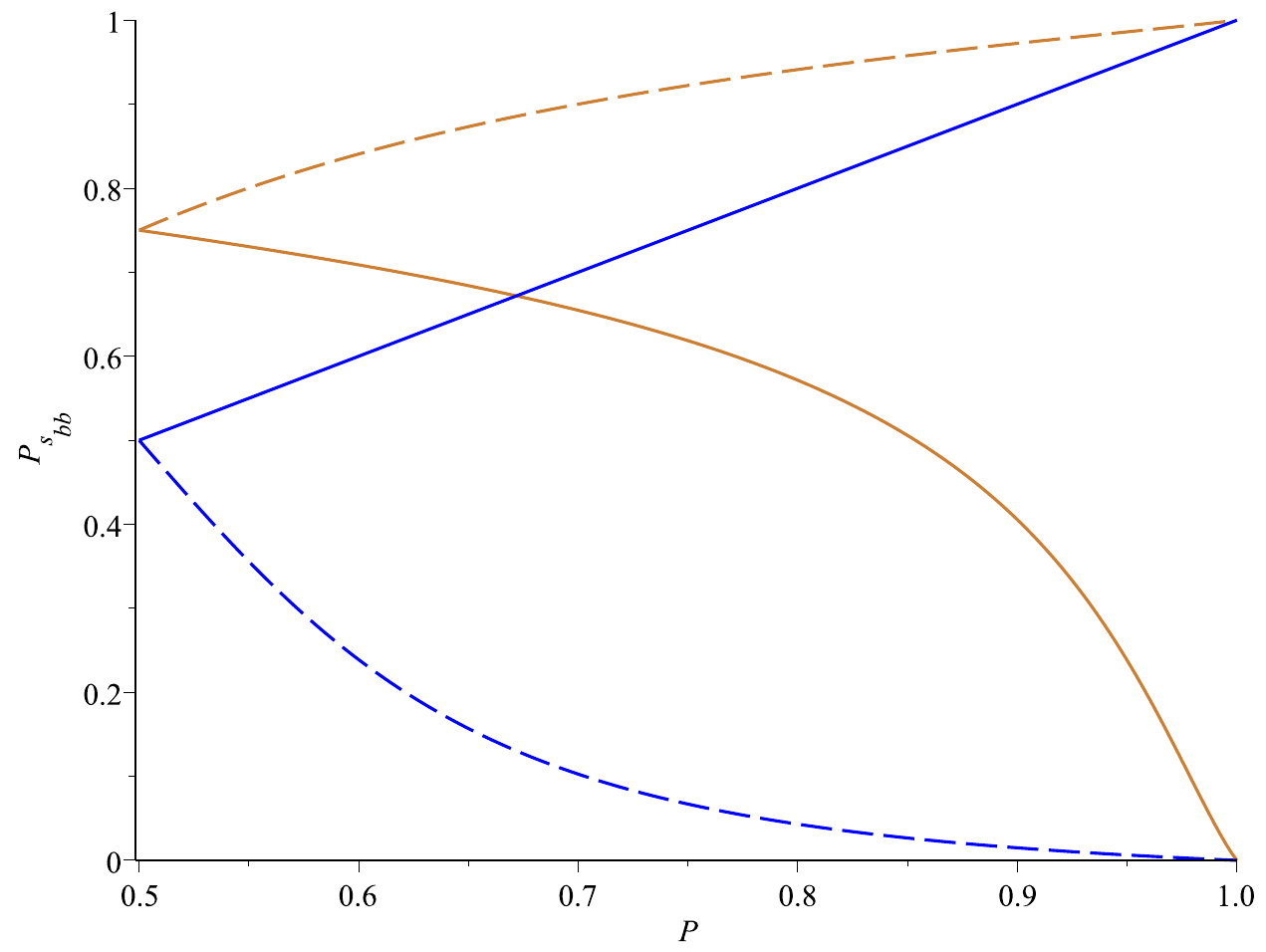}
\caption{(Color online)  Prediction of the heavy-quark spin-content
parameters $P_{s_{b\bar b} = 1}$ of Eqs.~(\ref{eq:AllPValues}) for the
lighter of mass eigenstates that are degenerate in $J^{PC}$, as
functions of the parameter $P$ defined in Eq.~(\ref{p_1_pm}).  Solid
(dashed) lines indicate $I \! = \! 1$ ($I \! = \! 0$) states.  These
levels from top to bottom at $P \! = \!  3/4$ are: $\bar X_0^{I=0}$
(dashed gold); $\bar Z^{I=1}$ (solid blue, which is $P$ itself); $\bar
X_0^{I=1}$ (solid gold); $\bar Z^{I=0}$ (dashed blue).
\label{fig:PValues}
}
\end{figure}

The heavy-quark spin structure of the mixed eigenstates can also be
computed solely as functions of $P$, according to
Eqs.~(\ref{eq:AbbrevsWithP})--(\ref{eq:AllPValues}).  The results are
presented in Fig.~\ref{fig:PValues}.  We find in the range $1/2 \!
\leq \! P \! \leq \! 1$ that $\bar X_0^{I=0}$ decays preferentially
to $\Upsilon$ or $\chi_b$, $\bar Z^{I=0}$ to $h_b$ or $\eta_b$, and
the proportion for $\bar X_0^{I=1}$ depends sensitively upon the
precise value of $P$.

Having completed the analysis of the $\Sigma^+_g(1S)$ multiplet, we
now use the techniques of Ref.~\cite{Giron:2019bcs} to compute the
center of mass $M_0$ for any other multiplet.  The results of
Eqs.~(\ref{eq:bbParamValues}) indicate that $M_0(1S) \! = \!
10630$~MeV, with an uncertainty of no more than 5~MeV\@.  Using this
mass eigenvalue in a Schr{\" o}dinger equation with the
lattice-computed potentials of Refs.~\cite{Juge:1997nc,Juge:1999ie,
Juge:2002br,Morningstar:2019,Capitani:2018rox}, one finds the diquark
$\de \! \equiv \! (bq)_{\bar{\bf 3}}$ and its antiparticle $\bde \!
\equiv \! (\bar b \bar q)_{\bf 3}$ to have mass
\begin{equation}
m_\de=m_\bde=5383.1 \mbox{-} 5406.2 \;\rm{MeV} \, ,
\end{equation}
where the range indicates the effect of varying over potentials taken
from the different lattice simulations.  In turn, these $m_{\de,
\bde}$ values serve as inputs used to compute other multiplet
mass eigenvalues, and we predict
\begin{eqnarray}
M_0(1P)&=& 10960.9 \mbox{-} 10966.3\;\rm{MeV},\nonumber\\
M_0(2S)&=& 11087.7 \mbox{-} 11093.2\;\rm{MeV}.
\end{eqnarray}
One immediately notes that of the two remaining $b\bar b q\bar
q^\prime$ candidates in Table~\ref{tab:bbqqExpt} (both with negative
parity), $Y(10860)$ lies about 70~MeV below $M_0(1P)$ and thus
uncontroversially fits into the $1P$ multiplet.\footnote{In
comparison, the lowest $1^{--}$ state $Y(4230)$ in the $c\bar c q\bar
q^\prime$ $\Sigma^+_g(1P)$ multiplet lies about 140~MeV below the
multiplet center of mass and yet fits well in the
multiplet~\cite{Giron:2020fvd}.}  On the other hand, $Y(10750)$ does
not fit well into this scheme; indeed, it is only about 100~MeV
heavier than the $1S$ state $Z_b(10650)$.  The $nP$-$nS$ average mass
splitting for conventional bottomonium is about 450~MeV for $n \! = \!
1$ and 250~MeV for $n \! = \! 2$, suggesting that $Y(10750)$ is not
sufficiently heavy to be a $\Sigma^+_g(1P)$ $b\bar b q\bar q^\prime$
state.  However, it was noted even in the discovery
paper~\cite{Abdesselam:2019gth} that $Y(10750)$ is a natural candidate
for a higher conventional $\Upsilon$ state, likely identifying with a
missing $\Upsilon(nD)$ state, and possibly mixing with $\Upsilon(nS)$
states, although the exact composition remains a matter of
debate~\cite{Li:2019qsg,Chen:2019uzm}.  In support of this view, note
from Table~\ref{tab:bbqqExpt} that only $Y(10750) \! \to \! \Upsilon$
(but not $h_b$) decay modes have been observed to date, thus promoting
the hypothesis of a pure $s_{b\bar b} \! = \! 1$ state, as expected
for conventional bottomonium.

\subsection{Hidden-Charm/Strange Exotics}

As noted in the Introduction, the $c\bar c s\bar s$ sector was first
considered using a model with separated ($cs$) and ($\bar c \bar s$)
diquarks in Ref.~\cite{Lebed:2016yvr}.  The possibility that the
lightest $c\bar c s\bar s$ state is $X(3915)$ was introduced in that
work, a reprise of the arguments in favor of this assignment appearing
in Sec.~\ref{sec:ccss_ExptReview}.  We also noted that two strong
candidates for the sole $1^{++}$ state in the $c\bar c s\bar s$
$\Sigma^+_g(1S)$ multiplet, $Y(4140)$ and $Y(4274)$, have been
experimentally confirmed (Table~\ref{tab:ccssExpt}), but also that the
conventional charmonium state $\chi_{c1}(3P)$ is predicted to have a
mass and a width comparable to those observed for the two candidates.
Indeed, the early calculation of Ref.~\cite{Barnes:2005pb} predicts
\begin{equation}
m_{\chi_{c1}(3P)} = 4271 \; {\rm MeV}, \ \ \Gamma_{\chi_{c1}(3P)} = 39
\; {\rm MeV} \, .
\end{equation}

The Hamiltonian introduced in Ref.~\cite{Lebed:2016yvr} restricted to
the $\Sigma^+_g(1S)$ multiplet is actually identical to the one given
in Eq.~(\ref{eq:ccssFullHam}).  In Ref.~\cite{Lebed:2016yvr} it was
introduced as a purely phenomenological construct, but in this work it
is seen to be the direct expression of the dynamical diquark model,
and mass splittings between different BO multiplets can be computed
using lattice-calculated potentials, as in Ref.~\cite{Giron:2019bcs}.

A nagging difficulty with the $X(3915)$ has been an ambiguity in its
measured $J^{PC}$ quantum numbers.  As suggested in
Table~\ref{tab:ccssExpt} and discussed by the
PDG~\cite{Tanabashi:2018oca}, the original $0^{++}$ assignment relies
on the assumption of dominance by a particular $\gamma \gamma$
helicity component in $X(3915)$ production, and if this assumption is
relaxed then the assignment $2^{++}$ is also possible.

Using the measured masses in Table~\ref{tab:ccssExpt}, we therefore
obtain fits to the Hamiltonian of Eq.~(\ref{eq:ccssFullHam}) under two
alternate assumptions: that the $X(3915)$ is the lighter of the two
$0^{++}$ states in $\Sigma^+_g(1S)$, or that it is the sole $2^{++}$
state.  For the moment we also assign $Y(4140)$ to be the sole
$1^{++}$ state, supposing by default that $Y(4274)$ is
$\chi_{c1}(3P)$.  The results of fits with both $X(3915)$ assignments
are presented in Table~\ref{tab:ccss_mass_predictions}.  In either
case, the spectrum is quite simple, consisting of only 3 distinct (and
equally spaced) mass eigenvalues for the 6 states.

\begin{table}[h]
\centering
\caption{Prediction of the 6 state masses (in MeV) of the
$\Sigma^+_g(1S)$ $c\bar c s\bar s$ multiplet, using the Hamiltonian of
Eq.~(\ref{eq:ccssFullHam}).  Boldface indicates the measured $X(3915)$
and $Y(4140)$ masses used as inputs for the fit.}
\setlength{\extrarowheight}{1.0ex}
\begin{tabular}{c c c | c c }
\hline\hline
$J^{PC}$ & \multicolumn{2}{c|}{$J_{X(3915)}^{PC}=0^{++}$} &
\multicolumn{2}{c}{$J_{X(3915)}^{PC}=2^{++}$} \\[0.5ex]
\hline
$0^{++}$& $\mathbf{3918.4}$ & $4375.2$ & $4375.2$ & $3918.4$ \\
$1^{+-}$& $4146.8$ & $4375.2$ & $4375.2$ & $3918.4$ \\
$1^{++}$& $\mathbf{4146.8}$ & $ $ & $\mathbf{4146.8}$ & $ $ \\
$2^{++}$& $4375.2$ & $ $ & $\mathbf{3918.4}$ & $ $ \\
\hline\hline
\end{tabular}
\label{tab:ccss_mass_predictions}
\end{table}

A stunning feature of Table~\ref{tab:ccss_mass_predictions} is that
both assignments predict a $0^{++}$ state at the mass of the
$X(3915)$, which suggests one remarkable scenario in which the
observed $X(3915)$ is actually a mixture of $0^{++}$ and $2^{++}$
states.  Furthermore, the third distinct mass in either case,
4375.2~MeV, lies quite close to that of the $X(4350)$, another $c\bar
c s\bar s$ candidate in Table~\ref{tab:ccss_mass_predictions}.
Confirmation of this state and a precise measurement of its mass and
$J^P$ quantum numbers ($C \! = \! +$ is known) at Belle~II will be
quite incisive.

These two fits, however, have a major difference that selects one as
more relevant to the spirit of the dynamical diquark model.  If
$X(3915)$ is $0^{++}$, then one obtains
\begin{equation}
M_0 = 4261.0\; {\rm MeV}, \ \ \kqQcs = +114.2\; {\rm MeV} \, ,
\label{parameters_pos_kappa}
\end{equation}
while taking $X(3915)$ to be $2^{++}$ gives
\begin{eqnarray}
M_0 &=& 4032.6\; {\rm MeV}, \ \ \kqQcs = -114.2\; {\rm MeV}.
\label{parameters_neg_kappa}
\end{eqnarray}
We have already noted in Sec.~\ref{sec:bbqqFits} that the diquark
spin-spin coupling $\kappa_{qQ}$ is positive in virtually every model,
so the scenario of Eq.~(\ref{parameters_neg_kappa}) leading to a
large, negative value of $\kqQcs$ and the $X(3915)$ being a degenerate
$0^{++}$-$2^{++}$ combination seems phenomenologically less appealing.

The large value of $\kqQcs$ obtained in
Eq.~(\ref{parameters_pos_kappa}) as compared to $\kappa_{cq}$ in
Eq.~(\ref{eq:ccqqFit}) or $\kqQb$ in Eq.~(\ref{eq:bbParamValues}) (a
factor of 5-6) suggests that the lighter constituent of the diquark
$\de$ has a significantly greater influence on the size of the
spin-spin coupling within $\de$ than does the flavor of the heavy
quark.  One may argue that the $s$ quark, being much heavier than $u$
or $d$, has less Fermi motion and allows $\de$ to be substantially
more compact, thus enhancing the effects of spin couplings within
$\de$.  In the language of quark models, the equivalent spin-spin
operator would have an expectation value scaling as some inverse
power of the $\de$ size.

Turning now to the identity of the sole $1^{++}$ state, we consider
the alternate possibility that $Y(4274)$ is a $c\bar c s\bar s$ state
and $Y(4140)$ is $\chi_{c1}(3P)$.  Then the third distinct mass
eigenvalue in the fits of Table~\ref{tab:ccss_mass_predictions}
becomes 4629.6~MeV, a much higher value than in the previous fit, and
completely unsuitable for the $X(4350)$.

Using the methods of Ref.~\cite{Giron:2019bcs} and the inputs of
Eq.~(\ref{parameters_pos_kappa}) [taking $X(3915)$ as the unique
lightest state and $Y(4140)$ as the sole $1^{++}$ state in the $c\bar
c s\bar s$ $\Sigma^+_g(1S)$ multiplet], we obtain
\begin{equation}
m_\de=m_\bde=2063.7 \mbox{-} 2085.5\;{\rm MeV} \, ,
\end{equation}
and predict
\begin{eqnarray}
M_0(1P) & = & 4625.3 \mbox{-} 4628.8\;\rm{MeV} \, , \nonumber \\
M_0(2S) & = & 4814.9 \mbox{-} 4818.1\;\rm{MeV} \, .
\end{eqnarray}
In comparison with the remaining states of Table~\ref{tab:ccssExpt},
the $\Sigma^+_g(1P)$ multiplet center of mass lies extraordinarily
close to that of $Y(4626)$, while $X(4500)$ is somewhat light to serve
as a $\Sigma^+_g(2S)$ state [plausibly, it could even be the heavier
$\Sigma^+_g(1S)$ $0^{++}$ state], but $X(4700)$ works well as the
lighter $0^{++}$ state in the $\Sigma^+_g(2S)$ multiplet.  Had
$Y(4274)$ instead been used for these fits, the results would have
been hundreds of MeV higher, reinforcing the conclusion that $Y(4140)$
works much better as a $c\bar c s\bar s$ state and $Y(4274)$ as
$\chi_{c1}(3P)$.

\section{Conclusions}
\label{sec:Concl}

This paper expands upon the work of
Refs.~\cite{Giron:2019bcs,Giron:2019cfc,Giron:2020fvd} to incorporate
the hidden-bottom ($b\bar{b} q \bar q^\prime$) and
hidden-charm/strange ($c\bar{c}s\bar{s}$) sectors into the dynamical
diquark model, primarily (but not exclusively) for the states that lie
in their respective ground-state [$\Sigma^+_g(1S)$] multiplets.

Starting from a Hamiltonian with only 3 parameters (for $b\bar b q\bar
q^\prime$) or 2 parameters (for $c\bar c s\bar s$) that describes the
fine structure within each multiplet of the model, we obtain explicit,
closed-form expressions for all 12 $b\bar b q\bar q^\prime$
isomultiplet masses and all 6 $c\bar c s\bar s$ masses.

In the $b\bar b q\bar q^\prime$ sector, the masses of the $Z_b(10610)$
and $Z_b(10650)$ combined with their relative preferences to decay to
$\Upsilon$ or $h_b$ states are sufficient to highly constrain all
other masses and heavy-quark-spin decay-mode preferences in the
$\Sigma^+_g(1S)$ multiplet.  In particular, the lightest states carry
$J^{PC} \! = \! 0^{++}$ and lie only a few 10's of MeV above the
$B\bar B$ threshold, and thus may have observably small widths.  The
$1^{++}$ analogue of the $X(3872)$ is predicted to lie in an
especially constrained range (10598-10607~MeV), near the $B\bar B^*$
threshold.

In a redux of the $c\bar c q\bar q^\prime$ sector (following on
Ref.~\cite{Giron:2019cfc}), we find that the 3-parameter Hamiltonian
also predicts an isoscalar $0^{++}$ state that is lighter than
$X(3872)$, but with exactly the right mass to merge with the
conventional charmonium $\chi_{c0}(2P)$ candidate at 3860~MeV.
Moreover, the fit values of the diquark internal spin coupling $\kqQb$
in the $b\bar b q\bar q^\prime$ sector and $\kappa_{qc}$ in the $c\bar
c q\bar q^\prime$ sector are numerically equal, but both are much
smaller than $\kqQcs$ in the $c\bar c s\bar s$.  The isospin-dependent
couplings $V_0$ in the $b\bar b q\bar q^\prime$ and $c\bar c q\bar
q^\prime$ sectors are both positive, having the same sign as the
corresponding pion-exchange operator in hadronic physics.

Once the center of mass for the $\Sigma^+_g(1S)$ multiplet is
determined from this analysis, we use potentials calculated in lattice
simulations to compute the corresponding centers for higher
multiplets, such as $\Sigma^+_g(1P)$ and $\Sigma^+_g(2S)$.  We find
that $Y(10860)$ works well as a $b\bar b q\bar q^\prime$ $1P$ state
but $Y(10750)$ is too light, very likely being primarily a $D$-wave
conventional bottomonium state.

In the $c\bar c s\bar s$ sector, we find it possible to identify
$X(3915)$ as a $2^{++}$ state, but only if the diquark spin coupling
$\kqQcs$ has opposite sign to the positive one nearly universally
accepted.  Thus the assignment $J^{PC} \! = \! 0^{++}$ is much more
natural in the dynamical diquark model.  Additionally, $X(4350)$
emerges directly as a $c\bar c s\bar s$ state.  We also find that
$Y(4140)$ is much more likely the sole $1^{++}$ $\Sigma^+_g(1S)$
$c\bar c s\bar s$ state and $Y(4274)$ is the conventional charmonium
state $\chi_{c1}(3P)$.  Computing higher center-of-multiplet masses,
we find that $Y(4626)$ fits the $\Sigma^+_g(1P)$ multiplet well and
$X(4700)$ [but not $X(4500)$] fits the $\Sigma^+_g(2S)$ multiplet
well.

To summarize, the dynamical diquark model produces a large number
remarkable results, both in the fine structure of individual
multiplets by employing an extremely simple model, and in the
calculated splittings between multiplets, by using potentials
calculated from first principles on the lattice.  It further produces
interesting physical insights in multiple sectors of exotic states,
thus far including $c\bar c q\bar q^\prime$, $c\bar c s\bar s$, and
$b\bar b q\bar q^\prime$.  One could similarly analyze
hidden-charm/open-strange states, $B_c$-like exotics, pentaquarks, and
other possibilities.
\begin{acknowledgments}
This work was supported by the National Science Foundation (NSF) under 
Grant No.\ PHY-1803912.
\end{acknowledgments}

\bibliographystyle{apsrev4-1}
\bibliography{diquark}
\end{document}